\documentclass[a4paper,11pt]{article}
\pdfoutput=1 

\usepackage{jinstpub} % with a manual edit commenting out % \RequirePackage{amssymb}
\usepackage{subfiles}
\usepackage{caption}
\usepackage{subcaption}
\usepackage{braket}
\usepackage{natbib}
\usepackage[dvipsnames]{xcolor}

\usepackage{graphicx}
\graphicspath{ {./figures} }

\def\MJ{{\sc Majorana}}

\def\nonubb{$0\nu\beta\beta$ }
\def\ndbd{$0\nu\beta\beta$ decay }

\title{\boldmath CAGE: An Internal Source Scanning Cryostat for HPGe Characterization}

\author[a,b,1,2]{G.~Othman,\note{Corresponding authors.}\note{Current address: Munich Quantum Instruments GmbH, Lichtenbergstra{\ss}e 8, 85748 Garching, Germany}}
\author[c,1]{C.~Wiseman,}
\author[c]{T.H.~Burritt,}
\author[c]{J.A.~Detwiler,}
\author[d]{M.P.~Held,}
\author[a,b]{R.~Henning,}
\author[c]{T.~Mathew,}
\author[c]{D.~Peterson,}
\author[e]{W.~Pettus,}
\author[c]{G.~Song,}
\author[c]{and T.D.~Van Wechel}

\affiliation[a]{Department of Physics and Astronomy, University of North Carolina, Chapel Hill, NC 27514, USA}
\affiliation[b]{Triangle Universities Nuclear Laboratory, Durham, NC 27708, USA}
\affiliation[c]{Center for Experimental Nuclear Physics and Astrophysics, and Department of Physics, University of Washington, Seattle, WA 98195, USA}
\affiliation[d]{Karlsruher Institut f{\"u}r Technologie Campus Nord, Eggenstein-Leopoldshafen 76344, Germany}
\affiliation[e]{Center for Exploration of Energy and Matter, and Department of Physics, Indiana University, Bloomington, IN 47405, USA}

\emailAdd{gulden.othman@gmail.com}
\emailAdd{wisecg@uw.edu}

\abstract{

The success of current and future-generation neutrinoless double beta ($0\nu\beta\beta$) decay experiments relies on the ability to eliminate or reduce extraneous backgrounds. In addition to constructing experiments using radiopure materials and handling in underground laboratories, it is necessary to understand and reduce known backgrounds in data analysis. The Large Enriched Germanium Experiment for Neutrinoless double beta Decay (LEGEND) is searching for \nonubb decay using $^{76}$Ge-enriched high-purity germanium (HPGe) detectors submerged in an active liquid argon (LAr) veto. A significant background in LEGEND is surface events from shallowly-impinging radiation on detector surfaces. In this paper we introduce the Collimated Alphas, Gammas, and Electrons (CAGE) scanning system, an internal-source scanning vacuum cryostat, designed to perform studies of surface events on sensitive surfaces of HPGe in a surface-lab. CAGE features a collimated radionuclide source inside a movable infrared shield that is able to perform precision scans of detector surfaces by utilizing three independent motor stages for source positioning. This allows detailed studies of pulse shapes as a function of source position and incident angle, where defining features can be extracted and exploited for removing surface backgrounds in data analysis in LEGEND. In this paper, we describe CAGE and demonstrate its performance with a commissioning run with $^{241}$Am. 
The commissioning run was completed with the source at normal incidence, and we estimate a beam spot precision of 3.1~mm, which includes positioning uncertainties and the beam-spot size. Using the 59.5~keV gamma population from $^{241}$Am, we show that low-energy photon events near the passivated surface feature risetimes that increase with radial distance from the detector center. We suggest a specific metric that can be used to discriminate low-energy gamma backgrounds in LEGEND with the similar characteristics (e.g.~$^{210}$Pb).

}


\arxivnumber{to be added}

\begin{document}
\maketitle
\flushbottom

\section{Introduction and Background}
\label{sec:intro}

The observation of neutrinoless double beta ($0\nu\beta\beta$) decay would have profound impacts on our understanding of the universe~\cite{Agostini:2022zub, 0vbb-2019, Gomez-Cadenas:2023vca}. Its observation would immediately indicate lepton number violation, an ingredient of leading models that can explain the observed matter-antimatter asymmetry in our universe~\cite{fukugida}. Additionally, $0\nu\beta\beta$ decay is the only practical way to determine whether neutrinos are their own antiparticles, i.e.~that they are Majorana fermions. Determining the Majorana or Dirac nature of the neutrino is essential to our understanding of the mechanism by which neutrinos obtain mass, can help explain the smallness of the neutrino mass, and gives insight into the neutrino mass ordering. 

With such strong theoretical motivation, many experiments have been mounted to search for this process in a variety of isotopes~\cite{Agostini:2022zub}. High-Purity Germanium (HPGe) detectors are one promising technology by which to discover $0\nu\beta\beta$ decay in the isotope $^{76}$Ge because of their excellent energy resolution and their strong background rejection capabilities via pulse shape analysis. The advantages of searching for $0\nu\beta\beta$ decay in $^{76}$Ge-enriched HPGe detectors have been demonstrated in two complementary previous-generation experiments, the \textsc{Majorana Demonstrator}~\cite{Abgrall:2025tsj} and GERDA~\cite{gerda-ins}. Among their generation, the \textsc{Majorana Demonstrator} achieved the best energy resolution~\cite{mjd-final} and GERDA the lowest background~\cite{gerda-final} in the region of interest for $0\nu\beta\beta$ decay. 

The LEGEND experiment~\cite{legend-pcdr} combines the best technologies of its predecessors to develop a tonne-scale search for $0\nu\beta\beta$ decay in $^{76}$Ge, using a phased approach installed in a LAr bath.
The first phase of LEGEND, LEGEND-200, will operate up to 200~kg of $^{76}$Ge with a background goal of 0.5~cts/FWHM~t~yr, aiming to reach a half life discovery sensitivity of $10^{27}$~yr after 5 years of operation.
LEGEND-200 is now operating in the former GERDA infrastructure at LNGS, having released initial results based on 61~kg~yr of data~\cite{l200-results-arxiv}.
The subsequent LEGEND-1000 phase will scale up to an enriched mass of 1000~kg, targeting a background level of $<0.025$~cts/FWHM~t~yr and a half life discovery sensitivity of $>10^{28}$~yr~\cite{legend-pcdr}.

In LEGEND, the primary source of surface backgrounds within the 2039~keV region-of-interest (ROI) for \ndbd in $^{76}$Ge is expected to be shallowly impinging alpha and beta radiation. Alpha contamination originates from $^{210}$Po (a $^{238}$U decay chain isotope), and betas from $^{42}$K (daughter of $^{42}$Ar) in the LAr active veto. $^{210}$Po may plate-out on the detector surfaces where it later alpha-decays, while $^{42}$K ions in the liquid argon (LAr) active veto may drift due to the external electric field lines of the detectors and beta-decay near the surfaces. 
$^{210}$Po alphas have an energy of  5.3~MeV, and $^{42}$K betas have an endpoint of 3.5~MeV. Because of surface effects, alpha and beta events can become energy-degraded and be reconstructed within the 2039~keV ROI for \ndbd in $^{76}$Ge. Additional rare-event searches with LEGEND, such as dark matter searches, are also impacted by surface events outside the ROI for \nonubb decay.  For example, low-energy gammas quickly attenuate in the detector surfaces and become energy-degraded. Compton scatters from high-energy gammas can also occur near the surface of the detector and suffer from energy degradation, distorting the Compton background spectra in non-trivial ways. 

To maximize the discovery sensitivity for \ndbd of LEGEND-200 and LEGEND-1000, as well as additional rare-event searches, it is essential to understand and mitigate surface backgrounds.
In this paper we describe the Collimated Alpha Gamma and Electron (CAGE) test stand, a detector scanning system that was built to investigate surface events for the various HPGe detector geometries that will be used in the LEGEND experiment using movable collimated radionuclide sources in a vacuum cryostat. First, we summarize the origin of surface backgrounds and how they may affect the various detector geometries that will be employed in LEGEND. We then describe the design and subsystems of the test stand, including the alpha source collimator and motor system. We finish by showing the results of the first scans by CAGE using a $^{241}$Am alpha and gamma source.

\section{Point Contact (PC) High-Purity Germanium (HPGe) Detectors}
\label{sec:ppcs}

 Much of the success of \textsc{Majorana} and GERDA has been due to the development of Point Contact (PC) HPGe detectors \cite{Luke:1989oox,ppcs-barbeau}. The electrode geometry of PC detectors results in an extremely low capacitance ($\sim$pF) that reduces noise and improves energy resolution. The electrode geometry also results in long drift-times and fast charge-collection, enabling excellent Pulse-Shape Discrimination (PSD) capabilities such as the discrimination of background-like multi-site events (MSE) from signal-like single-site events (SSE) \cite{avse, gerda-psd}.

\begin{figure}
    \centering
    \includegraphics[width=\textwidth]{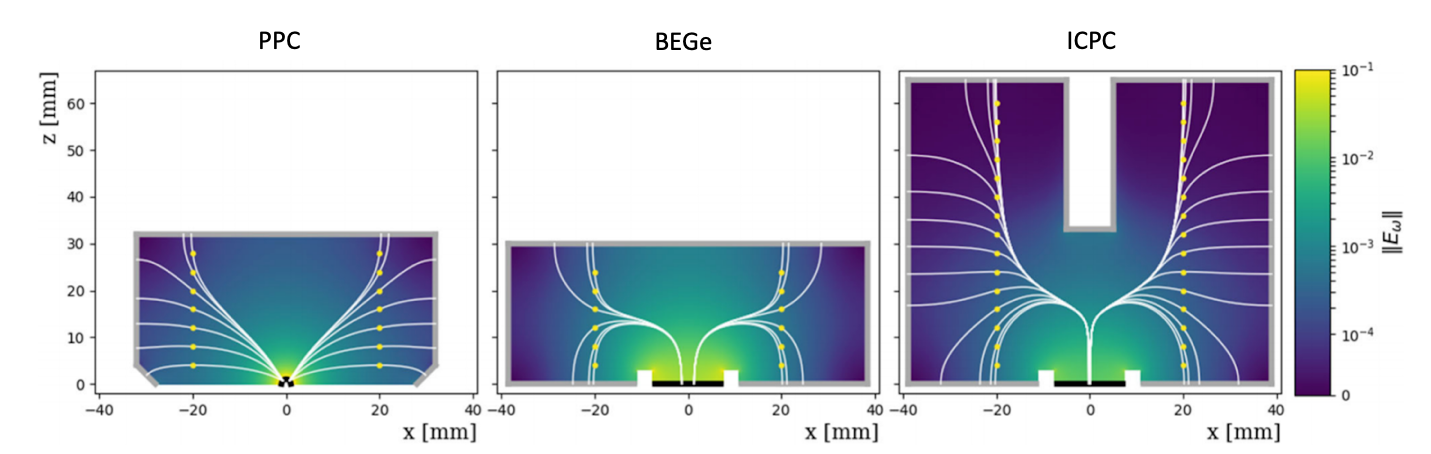}
    \caption{Two-dimensional cross-section of PC detector technologies used in the LEGEND-200 experiment, with weighting field magnitude, $|E_w|$, indicated by the colorbar. The heavy grey lines denote the N+ surfaces, the heavy black lines indicate the P+ contacts, and all surfaces with no heavy lines are bare or passivated. (left) \MJ~style PPC detector with large passivated surface; (center) GERDA Broad Energy Germanium (BEGe) detector, with a passivated ditch and large P+ contact; (right) Inverted Coaxial Point Contact (ICPC) detector, implemented for LEGEND.  Figure from~\cite{Comellato:2020ljj}.}
    \label{fig:ppcxs}
\end{figure}

 Generally, PC detectors operate as reverse-biased P-N junctions with a Li-diffused N+ outer contact ($\sim$1~mm thick) surrounding much of the surface of the crystal, plus a small ($\sim$1-10~mm diameter, $\sim$0.3~$\mu$m thickness) boron-implanted electron-blocking P+ contact. A thin passivation layer ($\sim$0.1~$\mu$m), often formed by amorphous-Ge (aGe) sputtering or silicon monoxide deposition, acts as an electrical insulator between the P+ contact, held near ground, and the N+ electrode, held at $\sim$kV high-voltage (HV). Ionizing radiation in the detector liberates holes and electrons, which drift along the electric field lines to the P+ and N+ contacts. The signal induced on the P+ contact is read out and digitized using dedicated electronics. The PC detector geometries  employed in LEGEND-200 are shown in Fig.~\ref{fig:ppcxs}. In addition to the PPCs and Broad Energy Germanium (BEGe) detectors used in \MJ ~and GERDA, LEGEND operates a new detector geometry called Inverted-Coaxial Point-Contact (ICPC) \cite{icpc, icpc-psd}. ICPC detectors combine the desired low-capacitance, long-drift-time, fast charge-collection features of PC detectors with the larger size of semi-coaxial HPGe detectors, and have been fabricated with masses up to 4~kg. 

With different deposition processes, composition, and thicknesses, each detector surface type is unique and has a different response and susceptibility to surface events. The thickness of the N+ layer makes it less susceptible to alpha and beta backgrounds, but it is not a true dead-layer. Ionization charge from interactions occurring within the N+ region may diffuse into the detector bulk, resulting in incomplete charge-collection and energy-degraded events \cite{slow-pulse, graham-thesis}. Passivated surfaces are particularly troublesome due to being very thin, only $\sim$0.1 $\mu$m, making them susceptible to both alpha and beta radiation. Additionally, it is possible that static surface charges can build up on the passivated surface, pulling field lines within the detector to the surface, where there is lower electron and hole mobility~\cite{kevin}. Moreover, diffusion and self-repulsion impact charge drift and collection near surfaces, and there may be additional charge-trapping that is thought to be caused by surface deformities originating from the passivation method during manufacturing~\cite{deformities, canberra-deformities}. The culmination of these effects is an effective dead-layer at the passivated surface, resulting in energy-degraded events. Due to their comparatively large passivation regions, PPC detectors are particularly sensitive to these effects. 

LEGEND-200 currently employs enriched PPC, BEGe, and ICPC detectors, whereas LEGEND-1000 will feature  300-400 enriched detectors exclusively of the ICPC geometry.  While bulk characterization of all detectors to be installed in LEGEND is accomplished using dedicated rapid-scanning setups, with the CAGE system we can more thoroughly study surface events using natural-abundance Ge detectors in a surface laboratory. We accomplish this using movable collimated radiation sources in a vacuum cryostat to precisely set both the position and, uniquely, the incidence angle of the beam on a detectors top surface, allowing for scans of its passivated surface and any N+ surfaces that may be on the same face. Rotating the incidence angle effectively changes the interaction depth probed, enabling us to study the interactions at varying depths at the same radial position. We will use CAGE to study the pulse shapes of surface events and to develop pulse-shape-based cuts to discriminate that can then be applied in LEGEND analyses. In the remainder of this paper, we describe the principle and components of the CAGE scanning system, and present the results of first proof-of-concept commissioning runs with the collimator at normal incidence with respect to the detector surface.

\section{CAGE Overview and Systems}
\label{sec:cage_overview}

Alpha scanning has proven to be important in understanding surface interactions in HPGe detectors~\cite{tube,galatea,dcr,julieta-thesis, galatea-frank}. The Collimated Alphas, Gammas, and Electrons (CAGE) scanner was designed to meet specialized needs for detailed studies of particle interactions on passivated surfaces of HPGe detectors, including but not limited to alphas. It offers some advantages over previous scanning systems for HPGe detectors by including a spectral grade, collimated radiation source internal to a movable infrared (IR) shield, which is itself also able to rotate. Including the source within the infrared shield eliminates the need for slits in IR shield, which would allow IR-shine and lead to increased leakage current. With this design, we can position the source beam freely on the passivated surfaces of the (stationary) detector, also at varying incidence angles at the same radial position, which is unique to the CAGE system. 

\begin{figure}[ht]
    \centering
    \includegraphics[height=0.7\textwidth]{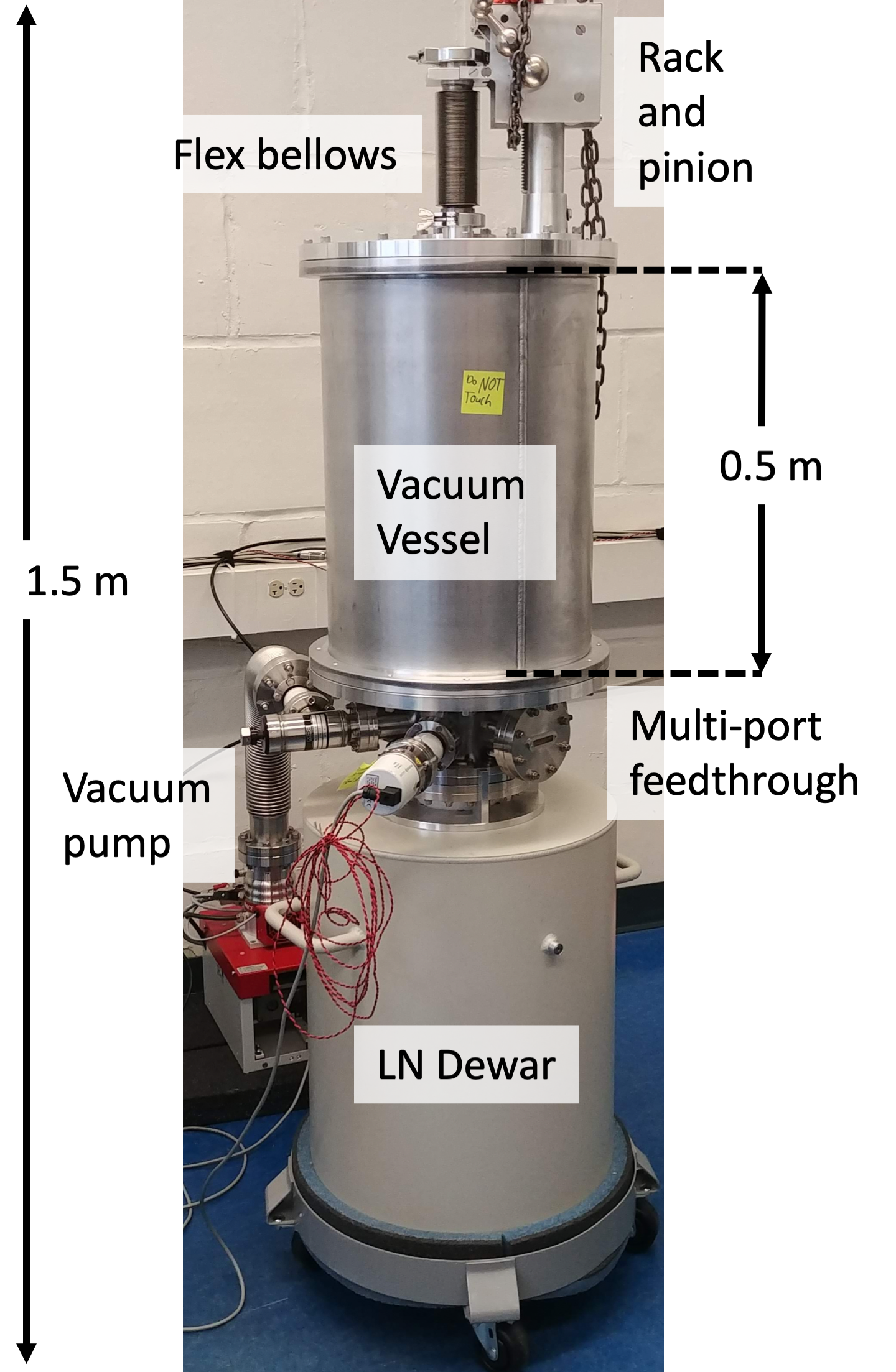}
    \caption{External view of the assembled CAGE vessel.}
    \label{fig:cage_vessel}
\end{figure}

The CAGE cryostat is a cylindrical vessel made of high-vacuum-grade aluminum, 0.5~m tall with a 0.35~m inner diameter, shown in Fig.~\ref{fig:cage_vessel}. The vessel is sealed by two aluminum lids with rubber O-rings. The bottom lid adapts the vessel to a multi-port conflat (CF) feedthrough flange below, which provides ports for the vacuum hose, signal readout, HV supply, and a pressure gauge. The lower portion of the multi-port feedthrough additionally provides the hardware for cooling the cryostat. It includes hardware on the vacuum side that accommodates a vacuum-side cold-finger extension, that is welded to a copper cold-finger on the air side. The multi-port feedthrough sits directly on a Liquid Nitrogen (LN) dewar, with an additional air-side copper cold-finger extension, and in this way provides LN cooling to the vessel. In the following sections, we describe the subsystems of CAGE in further detail. 

\subsection{Cryostat and Detector Mount}    
\label{sec:cage_vacuum}

\subsubsection{Detector Mount}
\label{sec:det_mount}

Inside the cryostat, a cylindrical copper cold plate 26.7~cm in diameter and 1.9~cm thick provides support and cooling for the detector and the infrared (IR) shield. 

\begin{figure}[ht]
    \centering
    \includegraphics[width=0.5\textwidth]{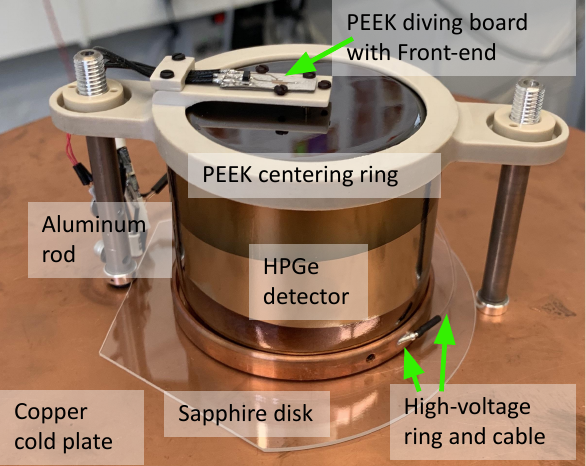}
    \caption{Detector mount on the CAGE cold-plate, shown here with the OPPI-1 PPC detector, described in \ref{sec:cage_collim}.}
    \label{fig:holder}
\end{figure}

\noindent As shown in Fig.~\ref{fig:holder}, the detector sits at the center of the cold-plate on a copper ring that supplies HV to the detector. We place a sapphire disk between the cold-plate and the HV ring. In this way, we electrically isolate the HV ring from the cold-plate while ensuring the detector can be sufficiently cooled.

Two aluminum rods thread into the cold plate on opposite sides of the detector, and allow a PEEK support ring to be lowered onto the detector surface. The PEEK ring ensures the detector is centered and secured on the cold-plate, and serves to hold the front-end electronics and contact pin. The front-end electronics sit on a ``diving board'' structure on the PEEK ring. Below the front end, a small hole in the diving board houses a spring-loaded contact pin that provides electrical contact between the front-end electronics and the P+ contact of the detector. The PEEK support structure was designed to be low-profile enough to ensure clearance with the collimator and source when moving them to different positions on the detector surface, while also providing enough standoff between the front-end and the N+ wraparound of ICPC detectors, which are held at HV.

\subsubsection{IR Shield and Motor Assembly}
\label{sec:ir-shield}

\begin{figure}[ht]
  \begin{subfigure}[t]{0.32\textwidth}
      \centering
      \includegraphics[scale=0.45]{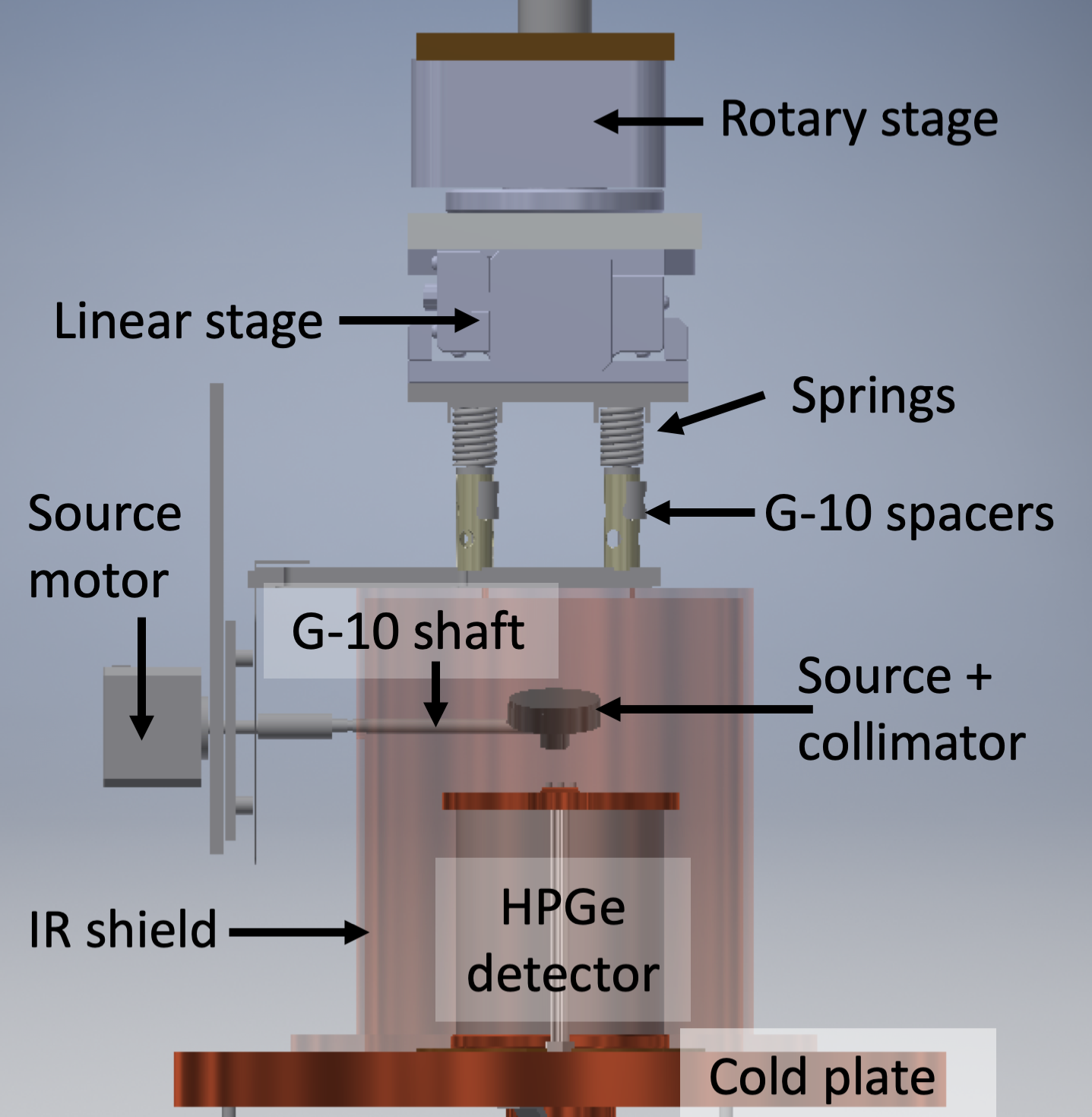}
    \caption{\label{subfig:labels}}
  \end{subfigure}
  \begin{subfigure}[t]{0.32\textwidth}
      \centering
      \includegraphics[scale=0.45]{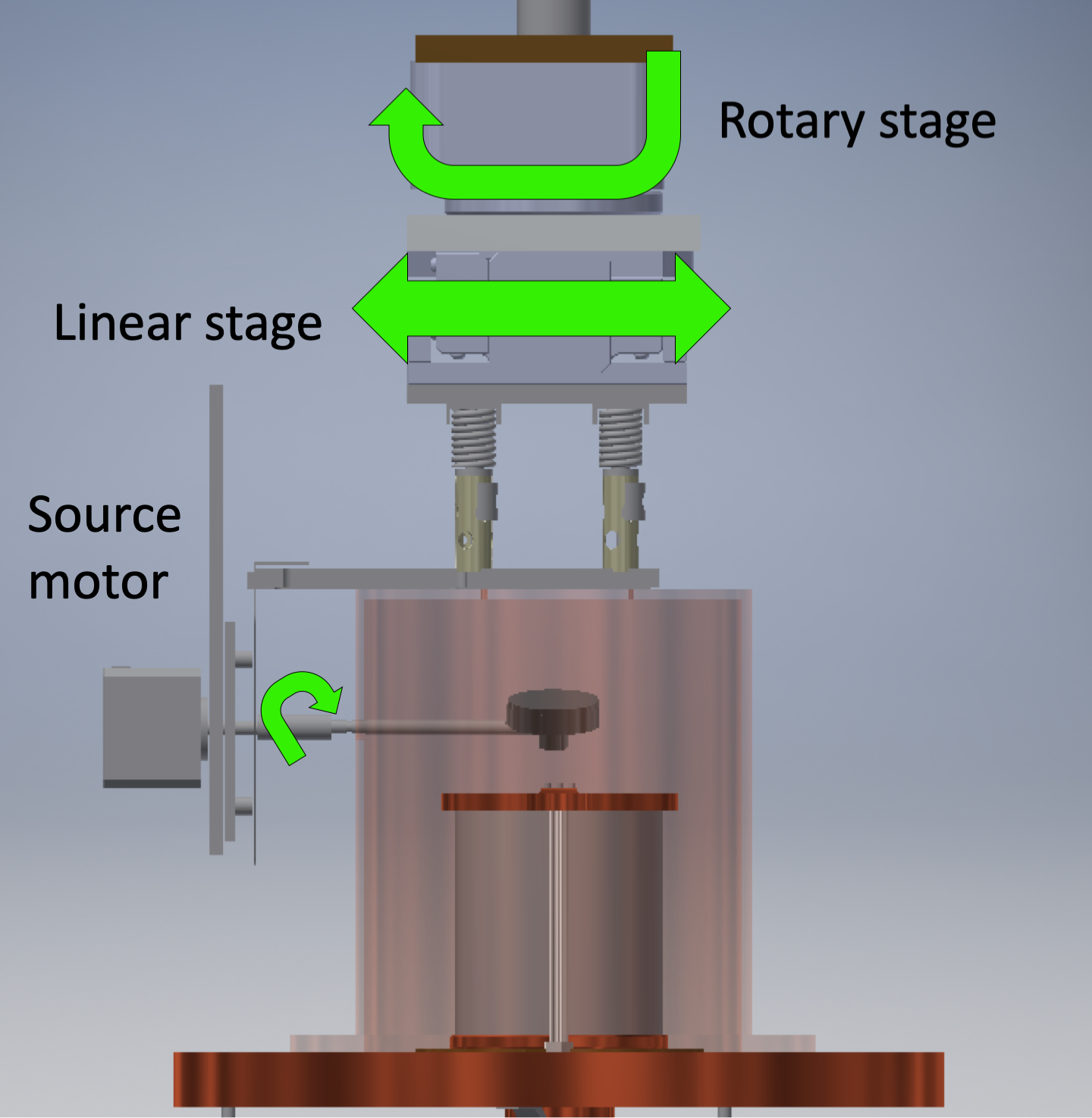}
    \caption{\label{subfig:movement}}
  \end{subfigure}
  \begin{subfigure}[t]{0.32\textwidth}
      \centering
      \includegraphics[scale=0.55]{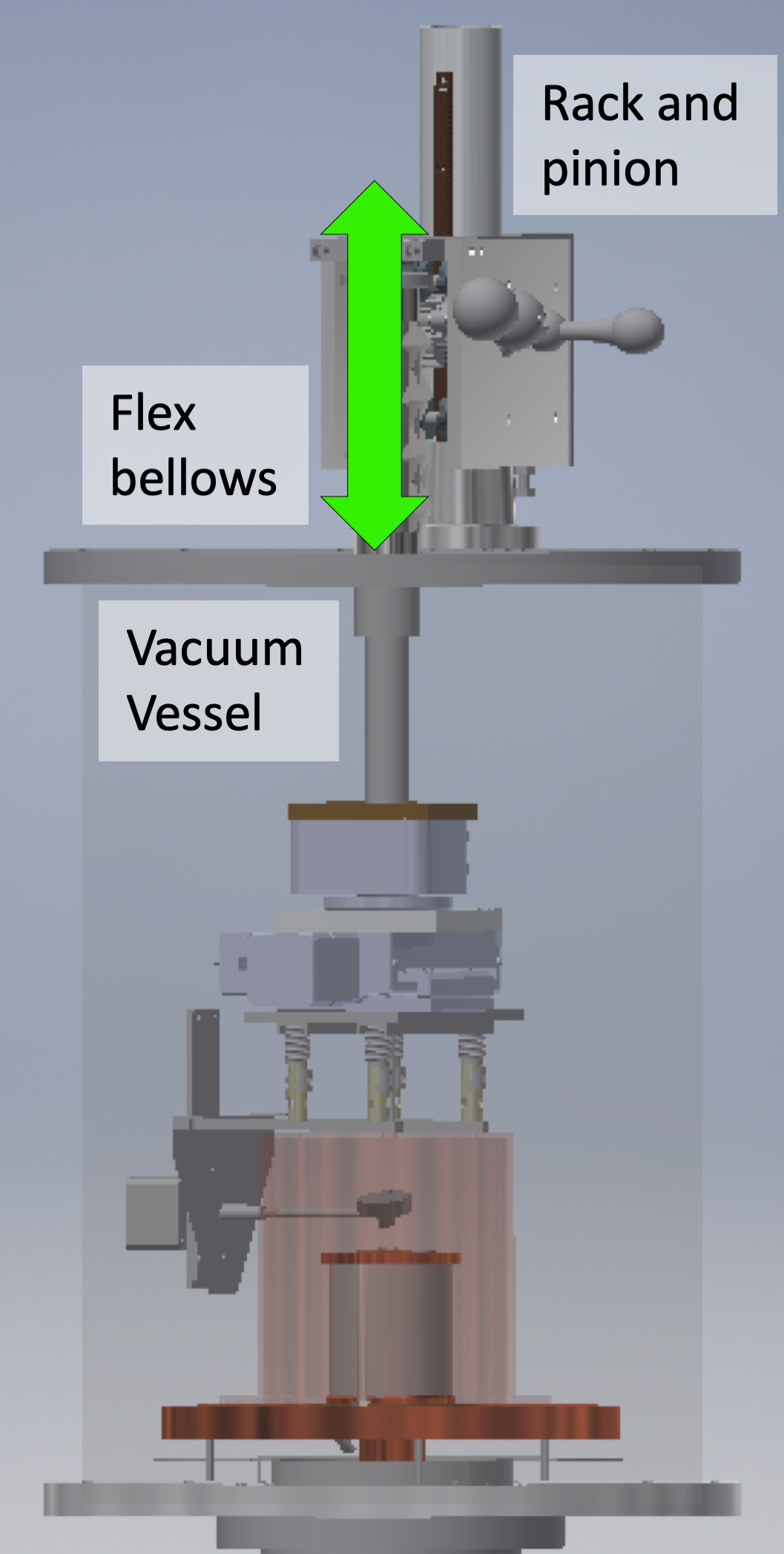}
    \caption{\label{subfig:pinion}}
  \end{subfigure}
  \caption{CAD rendering of the IR shield and motor assembly. (a) Shows the components, (b) shows the movement directions of the vacuum motors, and (c) shows the movement of the rack and pinion system to lift the IR shield assembly.  \label{fig:CAD_IR_assembly}}
\end{figure}

\begin{figure}[ht]
    \includegraphics[width=0.45\textwidth]{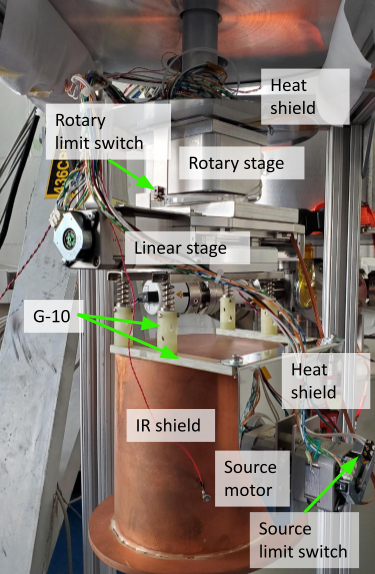}
    \hfill
    \includegraphics[width=0.45\textwidth]{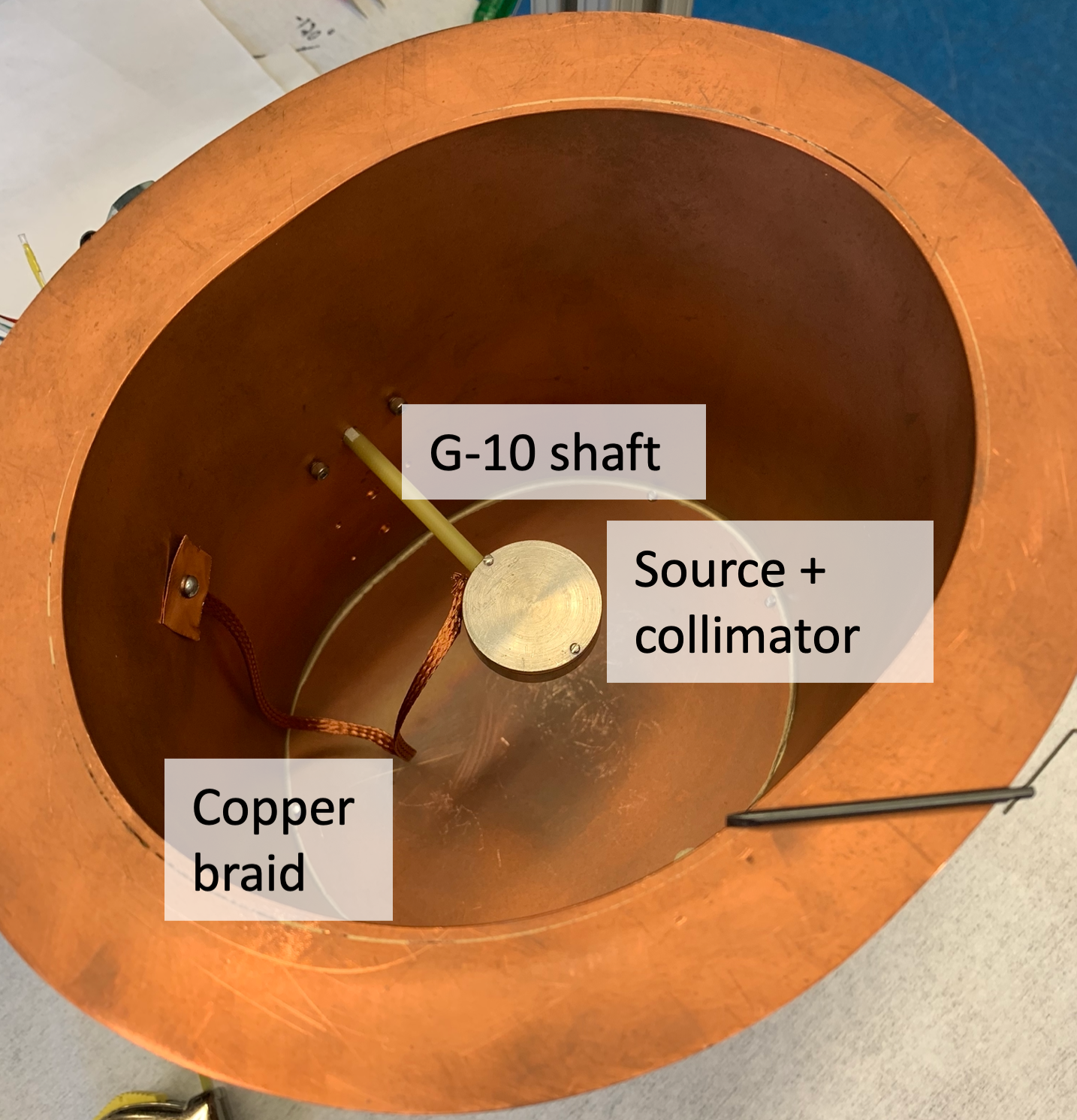}
    \caption{Left: IR shield with the motor assembly. The positions of the limit switches for the rotary stage and source motors are indicated here. The limit switches for the linear stage are internal and are not visible without disassembling the stage. Right: Underside of IR shield with G-10 shaft, collimator (pointed away from the camera), and copper braid.}
    \label{fig:tophat}
\end{figure}

The defining feature of CAGE is the ability to freely position the source on the detector surface while simultaneously protecting it from IR shine. We accomplished this by mounting a collimated radiation source directly onto and internal to a movable infrared (IR) shield. The IR-shield is a pill-box shaped copper tube resembling a ``top-hat'', with 2-mm-thick walls, a 16.7~cm major axis, and 13.3~cm minor axis. The ``brim'' of the IR top-hat is 2.5~cm wide. The dimensions of the IR-shield were chosen so that most ICPC and PPC detectors can fit inside with plenty of clearance. Two vacuum motor stages are mounted to the IR shield -- a rotary stage that rotates the entire shield 360 degrees, and a linear stage that translates the shield along a fixed axis. The position of the collimated beam on the detector surface is changed by lifting the entire IR shield and source assembly slightly, then moving it using these motor stages. The collimator itself is mounted within the IR shield on a third rotary stage, allowing the angle of incidence of the collimated source beam with respect to the detector surface to be changed. This design protects the passivated surface from IR shine while simultaneously allowing the source beam to be positioned freely on any point of the detector surface, with varying incidence angles from 45\textdegree\ to 90\textdegree\ (normal incidence). Varying the incidence angle would allow us to study depth dependent characteristics in addition to radial dependent ones, as it changes the effective interaction depth within the detector. Additionally, this movable IR shield design allows the detector to remain biased while changing the position of the source, which makes scans more fluid and consistent, as one does not have to un-bias and re-bias the detector between scan positions. For clarity, we show semi-transparent CAD renderings of the IR shield and motor stages for a detector installed on the cold-plate in Fig.~\ref{fig:CAD_IR_assembly}. We show a photo of the IR shield and the motor assembly, without a detector, in Fig.~\ref{fig:tophat} 

To reduce IR shine on sensitive passivated surfaces, the IR shield, source, and collimator must be cooled to near-LN$_2$ temperatures. This is accomplished by establishing good thermal contact with the cold-plate. However, while the IR shield is typically at a temperature of $\sim$90 K, the stepper motors employed in CAGE should be kept above 0\textdegree C to operate within their specifications. In order to meet this requirement, special care was taken in the thermal design to ensure an adequate temperature standoff between the IR shield and the motors in the assembly. This includes the use of thermally insulating G-10 components between the IR shield and motor stage, as well as stainless steel heat shields between the source motor stage to the side of the shield, and between the IR shield and the linear stage above the IR shield. These components are labeled in Fig.~\ref{fig:tophat}.

The collimator is directly mounted on a rotary stepper motor that sits outside the IR-shield, supported by two G-10 L-brackets from the top of the IR-shield. A G-10 shaft extension on the motor passes through the side of the IR-shield. Inside the IR shield, the collimator and source are attached directly to this G-10 shaft. To cool the source and collimator while leaving the motors warm, a copper braid thermally grounds the collimator to the IR-shield. The braid is of sufficient length to allow the collimator to be fully rotated. We show this in the photo in Fig. \ref{fig:tophat}.

Before the motors are used to move the IR-shield and position the source, the IR-shield must be lifted off the cold-plate to avoid scraping along the surface. For this purpose, the top lid of the aluminum vessel has additional hardware, including a rack and pinion assembly, to raise and lower the IR-shield and motor assembly from the cold-plate. A rod passes through the center of the top lid, supports the IR-shield assembly on the vacuum side by directly connecting to the upper rotary stage, and attaches to the rack and pinion on the air-side above. A KF flange with flex bellows allows the IR-shield to be manually raised and lowered onto the cold-plate using the rack and pinion while still maintaining high-vacuum inside the vessel. Four high-spring-constant springs  are installed between the G-10 spacers on the linear stage and the IR-shield. Compression of these springs when lowering the IR-shield onto the cold-plate helps ensure good thermal contact between the cold-plate and IR-shield.

\subsection{Movement System}
\label{sec:cage_motor}

\subsubsection{Movement Stage Requirements}

As mentioned in Sec.~\ref{sec:ir-shield}, CAGE uses three vacuum-rated movement stages to rotate and translate the IR shield and collimator, as shown in Fig.~\ref{fig:tophat}. We chose these stages based on stringent criteria. Since the cryostat maintains pressures below $10^{-7}$ mbar, motor stages inside the vessel must withstand ultra-high-vacuum and be able to be baked to reduce outgassing.
The stages should fit within the dimensions of the vacuum vessel while leaving enough space for at least 50~mm of linear travel, to accommodate the radius of most PPC detectors.
The entire detector surface must be easily accessible by the collimated beam through a combination of linear and rotary travel, and should achieve angles of incidence up to 45$^{\circ}$ relative to normal. Movement stages must be accurate enough to place the collimated beam within $\sim$1~mm of the desired location on the detector surface, with minimal hysteresis after repeated motions. Finally, since the stages are not visible from the outside, they must be capable of real-time in situ position calibration. We found that vacuum-rated Newmark motors and their corresponding controllers, coupled capacitive position encoders, meet our needs. We describe them in further detail in the following sections.

\subsubsection{Movement Stages}

A Newmark~RM-3 rotary positioner, which we refer to as the ``rotary stage,'' moves the IR shield about the central axis of the cryostat. A Newmark~NLS-4 linear positioner, or the ``linear stage,'' allows 5~cm of forward and backward movement from the center. Finally, a NEMA~17 motor, which we refer to as the ``source motor,'' changes the collimator angle and with it the angle of incidence of the source beam with respect to the detector surface. A Newmark~NSC motor controller powers and controls the stages, with movement routines and checks sent via Python to the controller. Each stage is moved by a NEMA~17 stepper motor with 1.8$^{\circ}$/step, and 250~microsteps/step resolution. The rotary stage has a worm gear drive with a 90:1 gear ratio, and the linear stage has a lead screw that translates at 1.5875~mm/rev, allowing for higher position resolution.

To read out the absolute position of the motors, we utilize a set of three 14-bit CUI AMT222B-V absolute position encoders, attached to the shaft of each motor. The encoders are single-turn, and are queried every 180$^{\circ}$ during movements to verify correct movement of each stage. 
To further determine the positions of the motors when installed in the cryostat, CAGE uses a set of limit switches that allow us to define ``zero'' positions of the motors and verify that the movement stages have returned to their zero positions, shown in Fig.~\ref{fig:tophat}. The source and rotary stages have a limit switch at 0$^{\circ}$, while the linear stage has a switch at each end of travel. Before positioning the source for data-taking, the stages are zeroed against the limit switches to avoid potential hysteresis effects. 

The movement stages are rated down to pressures of $10^{-7}$~mbar, which is limited by outgassing of low-vapor-pressure greases in the motors, and by the Teflon-insulated wires used to connect the stages to the controller. We find that operating pressures of 10$^{-8}$~mbar are achievable in the cryostat after baking, which is sufficiently low for HPGe detector operation. While the motor controller gives feedback on motor motions, operation at cold temperatures near the design limit can introduce additional friction in movements. We mitigate these effects using G-10 standoffs, as explained in Sec. \ref{sec:ir-shield}.

\subsubsection{Motor Accuracy and Precision Measurements}

To test the relative accuracy of the motor positioning system, we constructed an ``in-air'' test stand, with the IR shield and motor assembly supported by the aluminum frame visible in Fig. \ref{fig:tophat}. We moved the motors specified amounts by sending commands via software, and verified the resulting distances between the starting and ending positions with calipers and precision rulers. This allowed us to test the accuracy of the relative positions between motor movements. 

We attached a laser pointer to the IR shield and pointed at a flat surface to mark the position after movement.
We moved the linear stage in increments of 4~mm away from its limit switch, and then back again.
After repeated movements, we found that the linear stage has an uncertainty of $<$0.5 mm.
We tested the rotary and source stages in a similar manner by rotating in increments of 30$^{\circ}$, and found an uncertainty of $< 1^{\circ}$. These uncertainties were independent of the movement direction, and we find that hysteresis from the motor gearing is negligible at room temperatures.

In addition to the in-air tests, we conducted tests of the full assembly installed in the vessel under high-vacuum conditions. We drove all stages by set values away from their limit switches, and then back the same amount toward the limit switch. After multiple movements, each stage engaged its respective limit switch when predicted, and encoders confirmed the motors turned the desired amount successfully.

Though the motors themselves are extremely precise, it is more difficult to constrain the absolute positions of the entire IR-shield assembly, and thus the collimated beam, relative to the detector. We therefore refer to the beam positions in the remainder of the text as ``relative,'' as they are relative to the distance from the respective limit switches. During operation of the detector, it is possible that we can use the number of counts in the 60~keV line and the number of alpha events from the $^{241}$Am source to get an additional verification of the beam position from data via edge finding. For example, in Fig.~\ref{fig:60keV-rate} we see the rate of 60~keV gammas drop when the beam reaches the N+ surface of the detector, indicating an edge. This will be further investigated in future work.

\subsection{Electronics and High Voltage} 
\label{sec:cage_electronics}

We utilize a CAEN N1470 High-Voltage (HV) module to supply HV to the detector. The power is first filtered using a custom HV filter box that filters out high-frequency noise. The HV filter box is also designed with circuitry that allows us to send HV pulses to the detector, enabling checks of the full electronics chain. We route HV to the detector from the filter box via SHV cables on the air side of a vacuum feedthrough flange, and pico-coax cable on the vacuum side. These pico-coax cables are the same type that were used in \MJ, but we use a more robust connector since we are not concerned about radiopurity. The pico-coax cable is connected to the copper HV ring, the detector sits directly on the ring, and in this way HV is supplied (see Fig.~\ref{fig:holder}).

\begin{figure}[h!]
    \centering
    \includegraphics[width=0.6\textwidth]{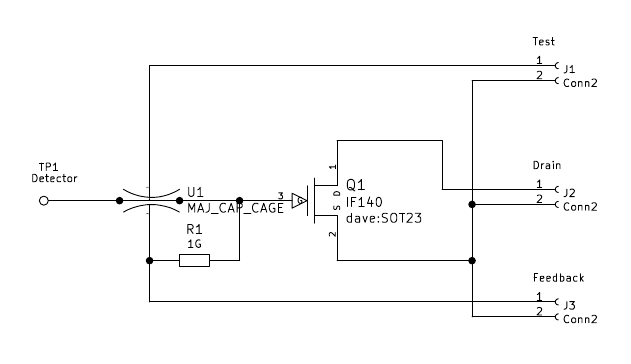}
    \caption{Schematic of the CAGE front-end board.}
    \label{fig:front-end}
\end{figure}

CAGE uses a set of custom amplifier circuitry designed at CENPA, similar to the amplifier chain used by the \MJ\ experiment~\cite{mjd_electronics}. The front-end preamplifier board (circuit schematic provided in Fig.~\ref{fig:front-end}; visible in Fig.~\ref{fig:holder}) is mounted on the PEEK detector holder, and the spring-loaded ``pogo'' pin connects the $p^+$ contact of the detector to the gate of an Interfet IF142 JFET.
To minimize electronics noise and maximize the bandwidth, the front-end board is mounted as close to the contact pin as possible.
The substrate material is Rogers 5880, which has low dielectric loss.
This choice makes the CAGE front-end board easier to fabricate than the fused silica boards used by \MJ, which was necessary for high radiopurity.
The front-end board has a resistor of 1~G$\Omega$ and feedback capacitance of 0.28~pF, giving pulses a characteristic decay time of $\sim$250~$\mu$s. The output signals from the front-end board are carried by a set of RG174 micro coax cables less than 1~m long to the signal feedthrough flange, and sent to a custom second-stage amplifier. The second-stage output, also designed at CENPA, is AC-coupled with a 50~$\mu$s RC time constant. The output of the amplifier is sent to a Struck SIS3302 100 MHz digitizer, housed on the same crate as HV module, and signals are recorded by ORCA, a real-time data acquisition software~\cite{howe2004}.

\subsection{Slow Controls Monitoring}
\label{sec:cage_scm}

The CAGE system has a number of slow controls sensors monitoring motor position (CUI AMT222B-V capacitive rotary encoders), pressure (Pfeiffer PKR 251), temperature (PT1000), detector baseline voltage (ADS1115) and high-voltage (CAEN N1470).
These sensors are connected to a Raspberry Pi (RPi) housed in a custom box.
To mitigate the possibility of ground loop noise from the sensors on the detector signal, the RPi is grounded to the outer cryostat body.
Some polling noise is present on the detector signal, but reducing the polling interval mitigates this effect.

To read and save values from passive sensors and to control active ones, we use the \texttt{dragonfly} slow controls monitoring tool~\cite{dragonfly}, originally developed by the Project 8 experiment~\cite{Project8} and based on the Dripline~\cite{dripline} framework.
Sensors connected to the system post messages at regular intervals, which are collected by a central server and stored in a PostgreSQL database, using a persistent Docker volume.

The high voltage card (CAEN N1470) can be controlled and polled remotely using \texttt{dragonfly}, which allows the detector to be biased and monitored remotely.
A software-based interlock for the HV system also monitors the baseline value over long periods of operation.
If the baseline suddenly deviates from its standard value (e.g.~due to high voltage breakdown), the interlock will automatically unbias the detector to prevent further damage, and send an alarm to operators.
Alarm notifications for anomalous slow control values or events are sent by \texttt{dragonfly} over email and Slack.
The ORCA data acquisition system is also set to send messages if the detector rate crosses a user-defined threshold.

\begin{figure}[ht]
    \centering
    \includegraphics[width=0.95\textwidth]{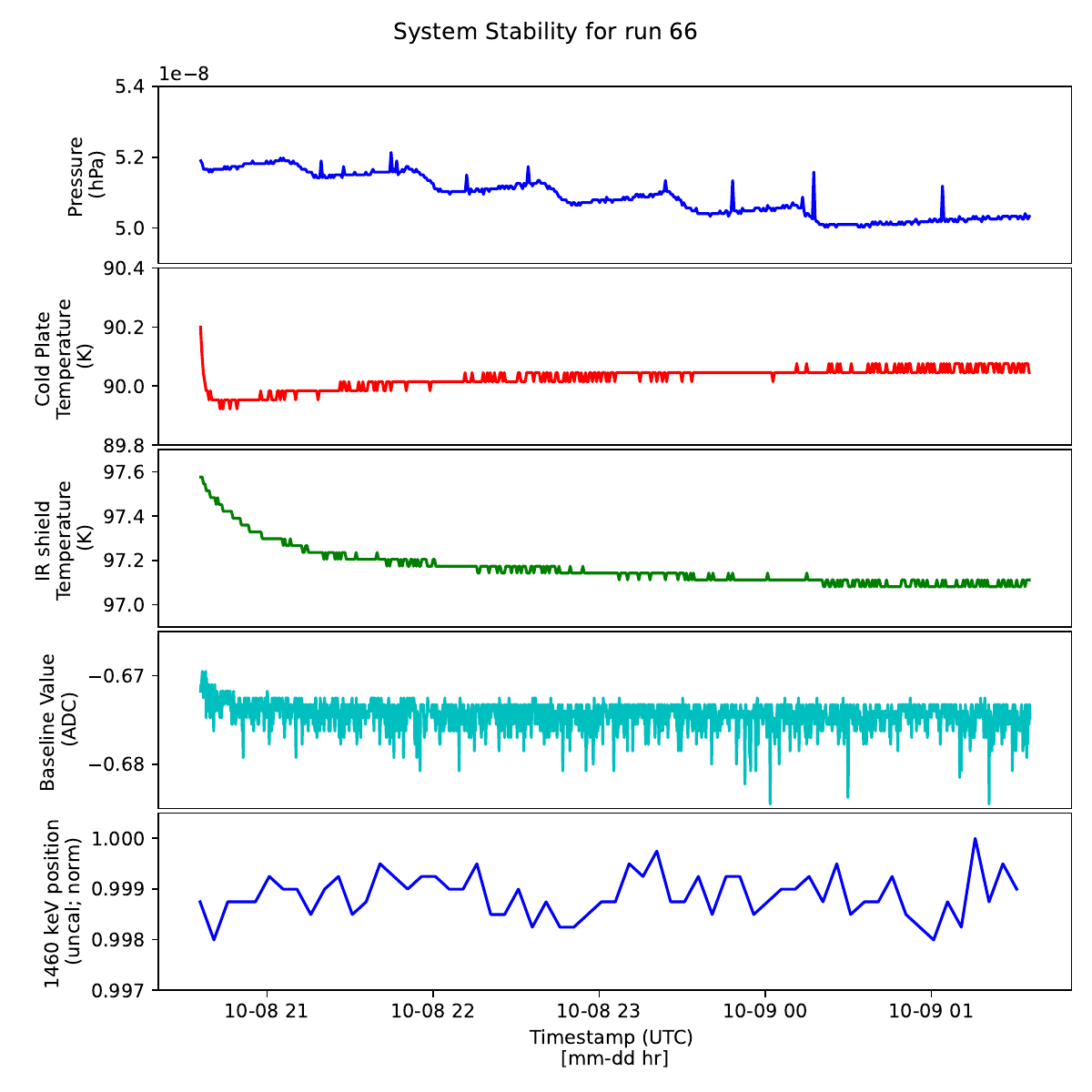}
    \caption{Top four panels: Slow controls metadata during data taking for a scan at 12.5 mm. Bottom panel: the uncalibrated ADC value of the 1460 keV gamma peak from $^{40}$K (from room background),
normalized so that the maximum value is 1, to show the stability of the gain of the detector throughout the run. The starting time corresponds to the lowering of IR-shield onto the cold plate. Spikes in the pressure are artifacts from the pressure sensor. The temperature of the cold plate is initially high when the IR-shield is still raised, because the IR shield is not protecting the temperature sensor from IR shine. After the shield is lowered, the temperature of the shield decreases and the temperature of the cold plate increases, until they each reach equilibrium around timestamp ``10-08 22.''}
    \label{fig:stability}
\end{figure}

The slow controls history provides useful insight into many aspects of the system's performance. A custom GUI allows convenient real-time readout of any combination of sensor values saved in the database, in addition to a Metabase~\cite{metabase} configuration which supports SQL queries to the database via a web interface for full history retrieval.  Figure~\ref{fig:stability} shows the vessel pressure, temperatures of the cold plate and IR shield, and the baseline voltage plotted against the position of the 1460~keV peak naturally present from $^{40}$K, during a 5 hour data-taking period. This provides verification of the system stability over the entire data-taking period, which begins after the cold plate and IR shield come into thermal equilibrium. 

\subsection{Collimator Design and Simulations} 
\label{sec:cage_collim}

\begin{figure*}[h!]
  \begin{subfigure}[t]{0.5\textwidth}
    \centering
      \includegraphics[scale=0.30]{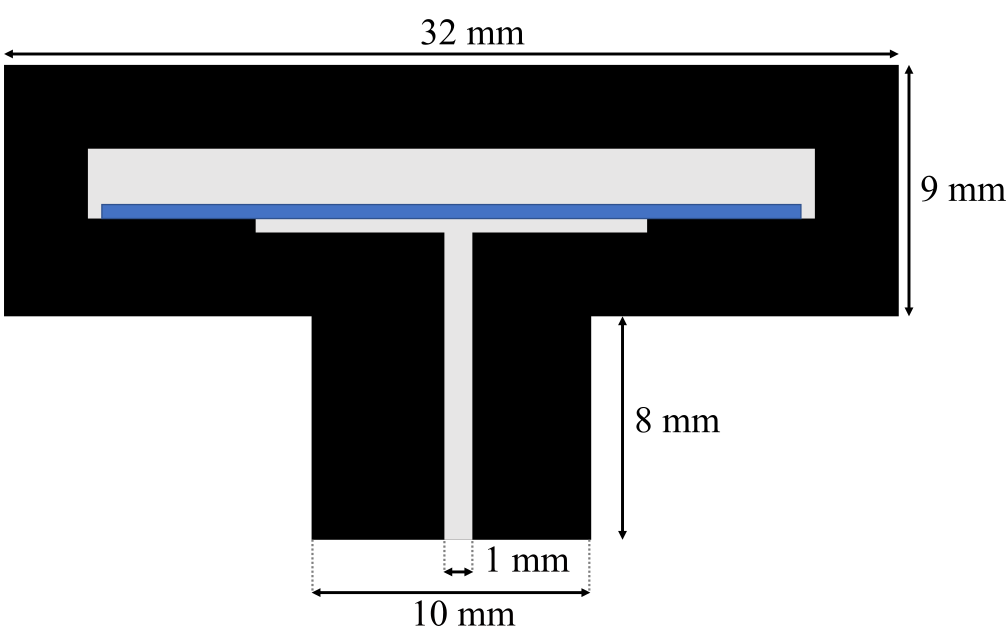}
    \caption{\label{subfig:coll_dims}}
  \end{subfigure}
  \hfill
  \begin{subfigure}[t]{0.45\textwidth}
    \centering
      \includegraphics[scale=0.18]{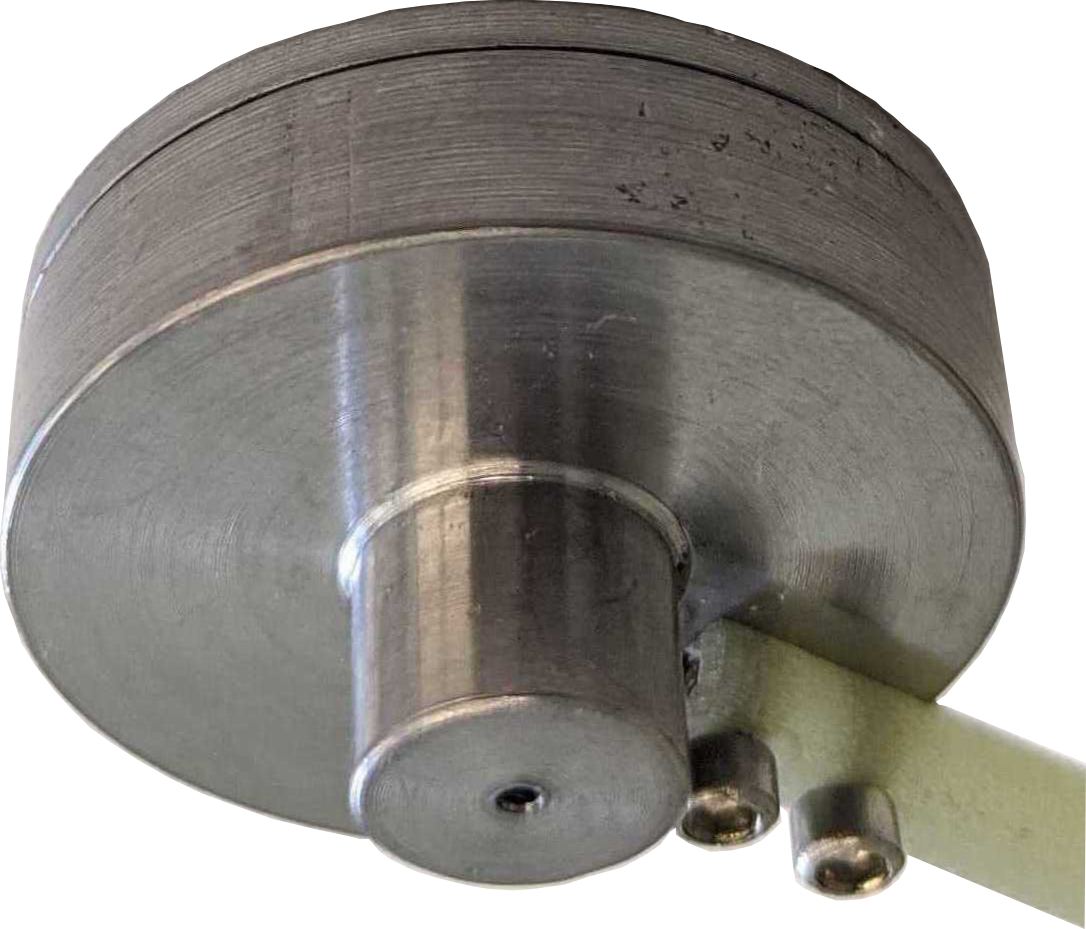}
    \caption{\label{subfig:coll_upright}}
  \end{subfigure}
\caption{(a) Cross-sectional drawing of the lead collimator. Light grey/white spaces indicate open areas. The blue rectangle represents the source disk. The activity is located on the inner 7~mm of the disk, slightly smaller in diameter than the grey/white cut-out below it.} (b) Photo of the lead collimator, including a portion of the G10 shaft for rotation.
\label{fig:collimator}
\end{figure*}

In the studies presented in this work, we utilize a commercially available spectral-grade $^{241}$Am source from  Eckert \& Ziegler~\cite{ez} with an activity of 40~kBq. We designed a custom collimator to use in conjunction with the source to precisely study the effects of radiation on specific locations on the detector surface.  In designing our collimator, we performed dedicated Monte-Carlo simulations based on g4simple \cite{g4simple} with an $^{241}$Am (alpha, 60~keV gamma) source within the collimator to ensure the collimator geometry provides a small beam spot-size while allowing a sufficient event rate for commercially-available source activities. Informed by these simulations, the collimator we designed for $^{241}$Am, shown in Fig. \ref{fig:collimator}, is made entirely from lead. The collimation hole is 1 mm, and the effective collimation length is 11.5 mm. The main body houses the source disk, and 3 mm thick walls ensure the alphas and gammas emitted from $^{241}$Am are sufficiently attenuated. Below the main body, there is an additional cylindrical section 8 mm long with a 1 mm hole for particle collimation. In the $^{241}$Am runs, we are interested in the 59.5 keV (which we call the 60 keV) gamma and 5.5 MeV alphas.

The collimator can be seen as-installed in the IR shield in Fig.~\ref{fig:tophat} (right). We have found that the mechanical design for the collimator and support, with the lead collimator cantilevered from one G-10 shaft, results in some uncertainty in the absolute beam position. In the results presented in this paper we use the collimator in this configuration (Fig.~\ref{fig:collimator}), while in future scans we will use an updated collimator design with both ends of the G-10 shaft supported, which limits the degrees of freedom of the collimator and therefore reduces uncertainty.

\subsubsection{Simulations Results with $^{241}$Am}
\label{sec:sims}

\begin{figure}[h]
\begin{center}
  \begin{subfigure}[t]{0.45\textwidth}
    \centering
      \includegraphics[scale=0.8]{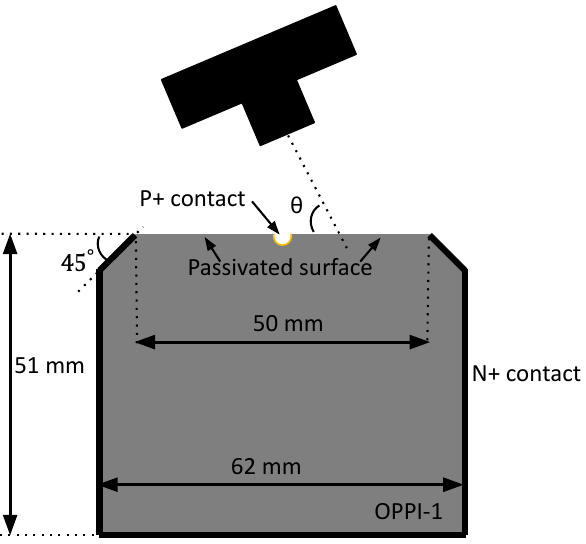}
    \caption{\label{subfig:oppi}}
  \end{subfigure}
  \hfill
  \begin{subfigure}[t]{0.45\textwidth}
    \centering
      \includegraphics[scale=0.33]{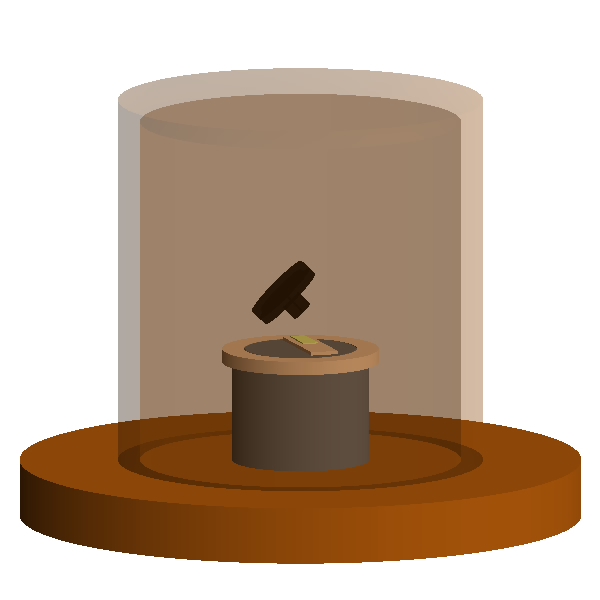}
    \caption{\label{subfig:sims_vis}}
  \end{subfigure}
\caption{Left: Cartoon (side-view) of the OPPI-1 PPC detector, used in simulations and as well as the data scans presented later. The N+ contact is indicated by the black solid line, and includes the bevel; it is made by Li implantation and is about 1~mm thick. The definition of the source angle is indicated by $\theta$. Right: Visualization of the simulated geometries from g4simple, with the source collimator (black) rotated at an arbitrary angle.}
\label{fig:oppi_sims_vis}
\end{center}
\end{figure}

\begin{figure}[h]
\begin{center}
  \begin{subfigure}[t]{0.45\textwidth}
    \centering
      \includegraphics[scale=0.30]{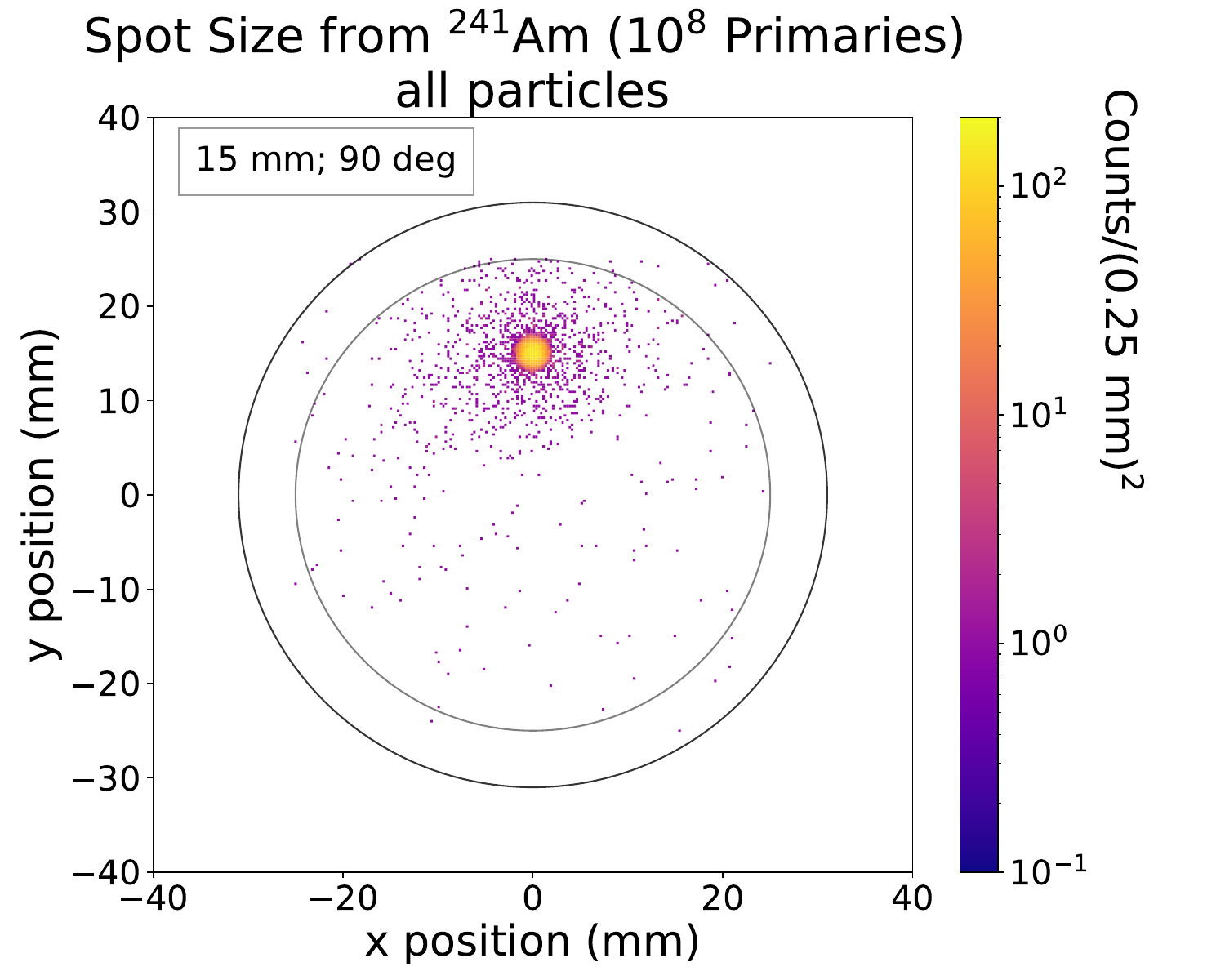}
    \caption{\label{subfig:2d_allParticles_90deg}}
  \end{subfigure}
  \hfill
  \begin{subfigure}[t]{0.45\textwidth}
    \centering
      \includegraphics[scale=0.30]{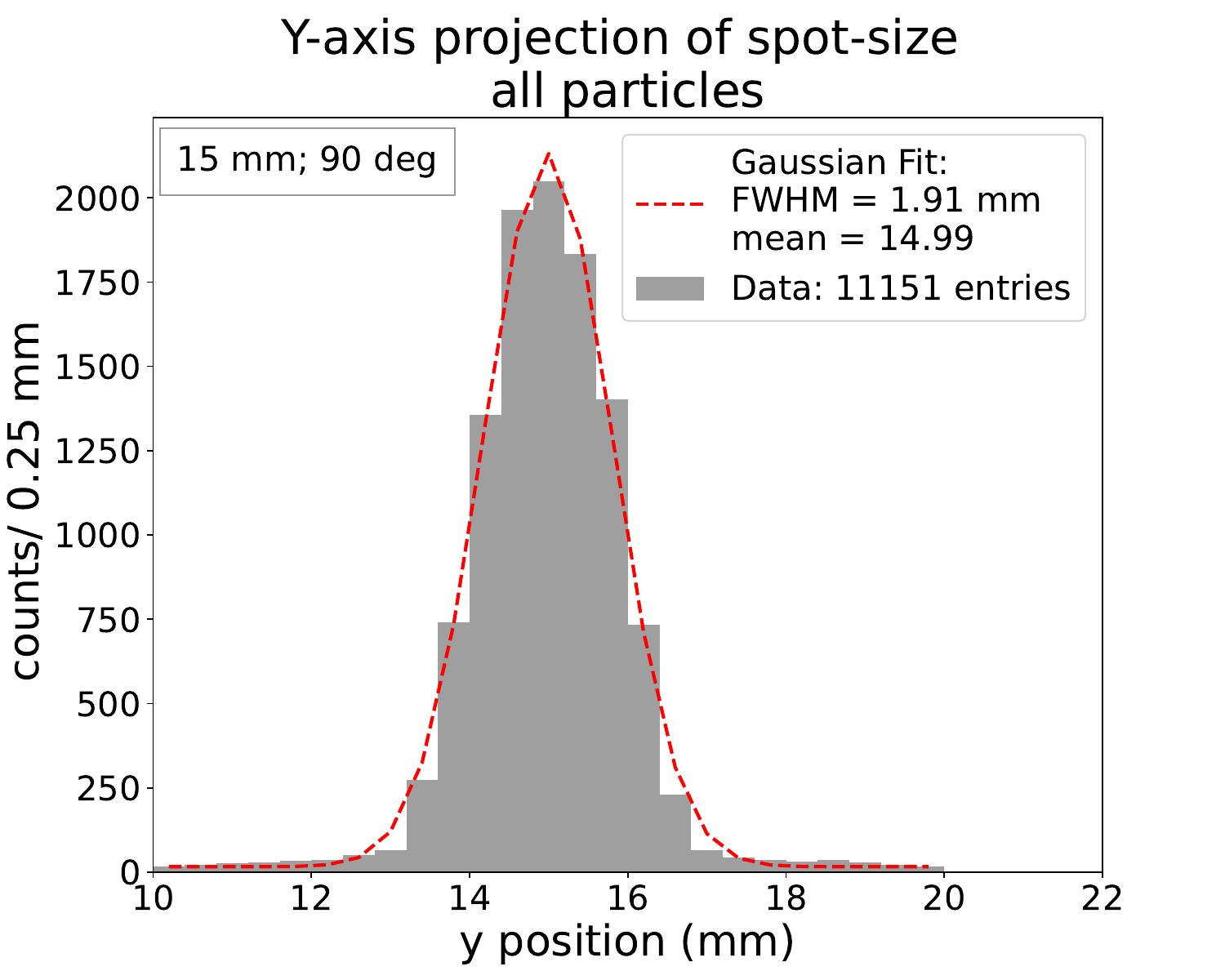}
    \caption{\label{subfig:1d_allParticles_90deg}}
  \end{subfigure}
  \begin{subfigure}[t]{0.45\textwidth}
    \centering
      \includegraphics[scale=0.30]{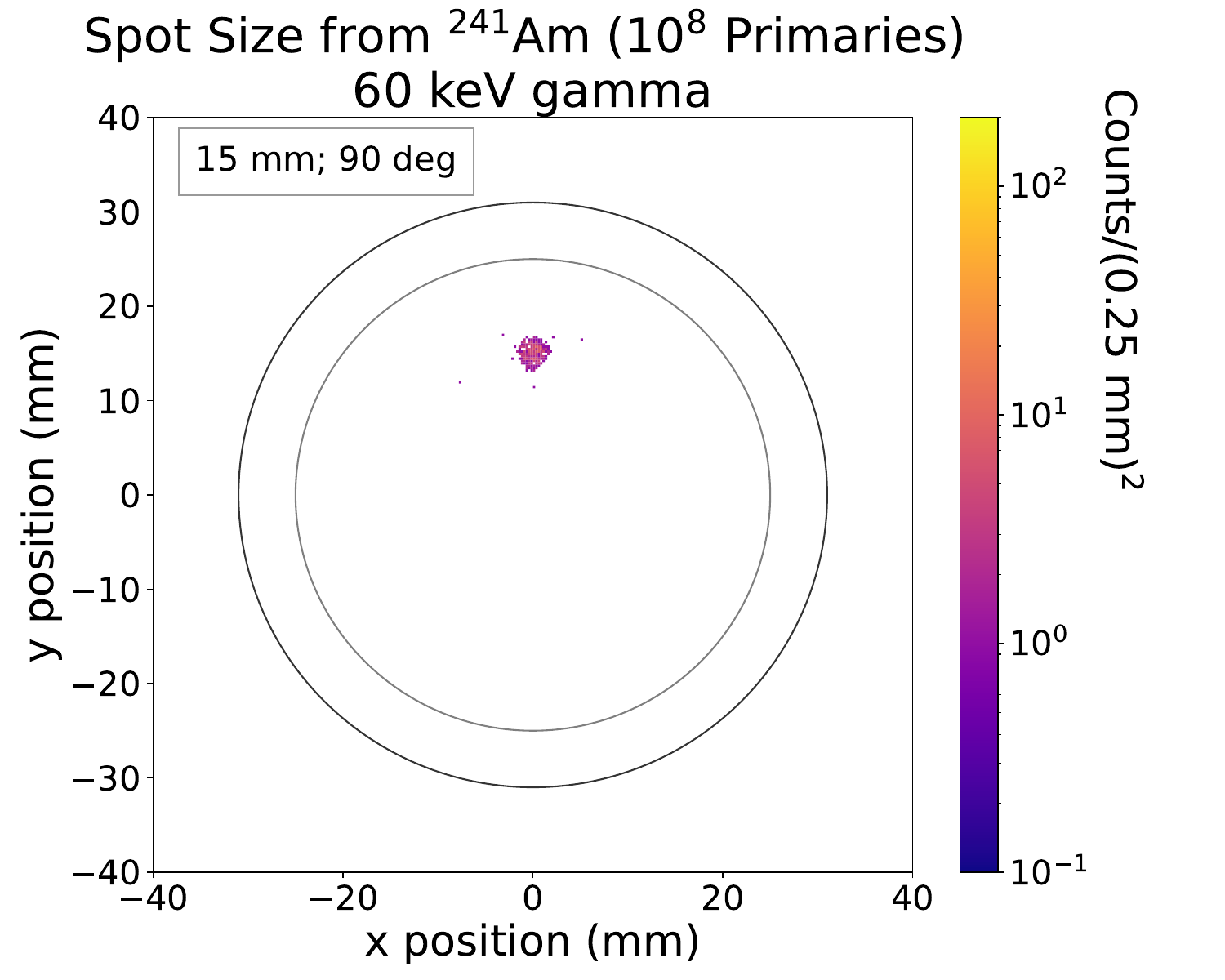}
    \caption{\label{subfig:2d_gammas_90deg}}
  \end{subfigure}
  \hfill
  \begin{subfigure}[t]{0.45\textwidth}
    \centering
      \includegraphics[scale=0.30]{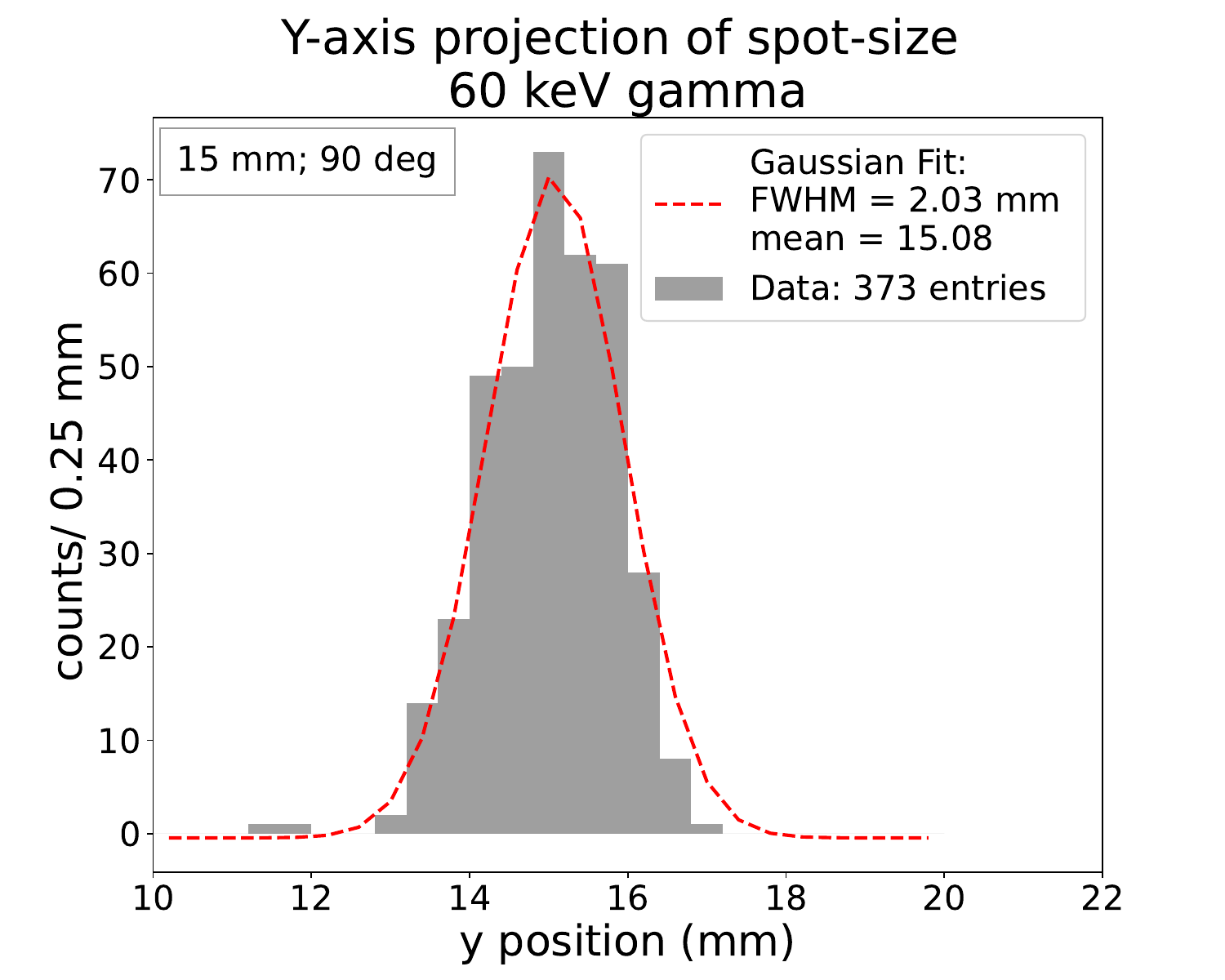}
    \caption{\label{subfig:1d_gammas_90deg}}
  \end{subfigure}
  \begin{subfigure}[t]{0.45\textwidth}
    \centering
      \includegraphics[scale=0.30]{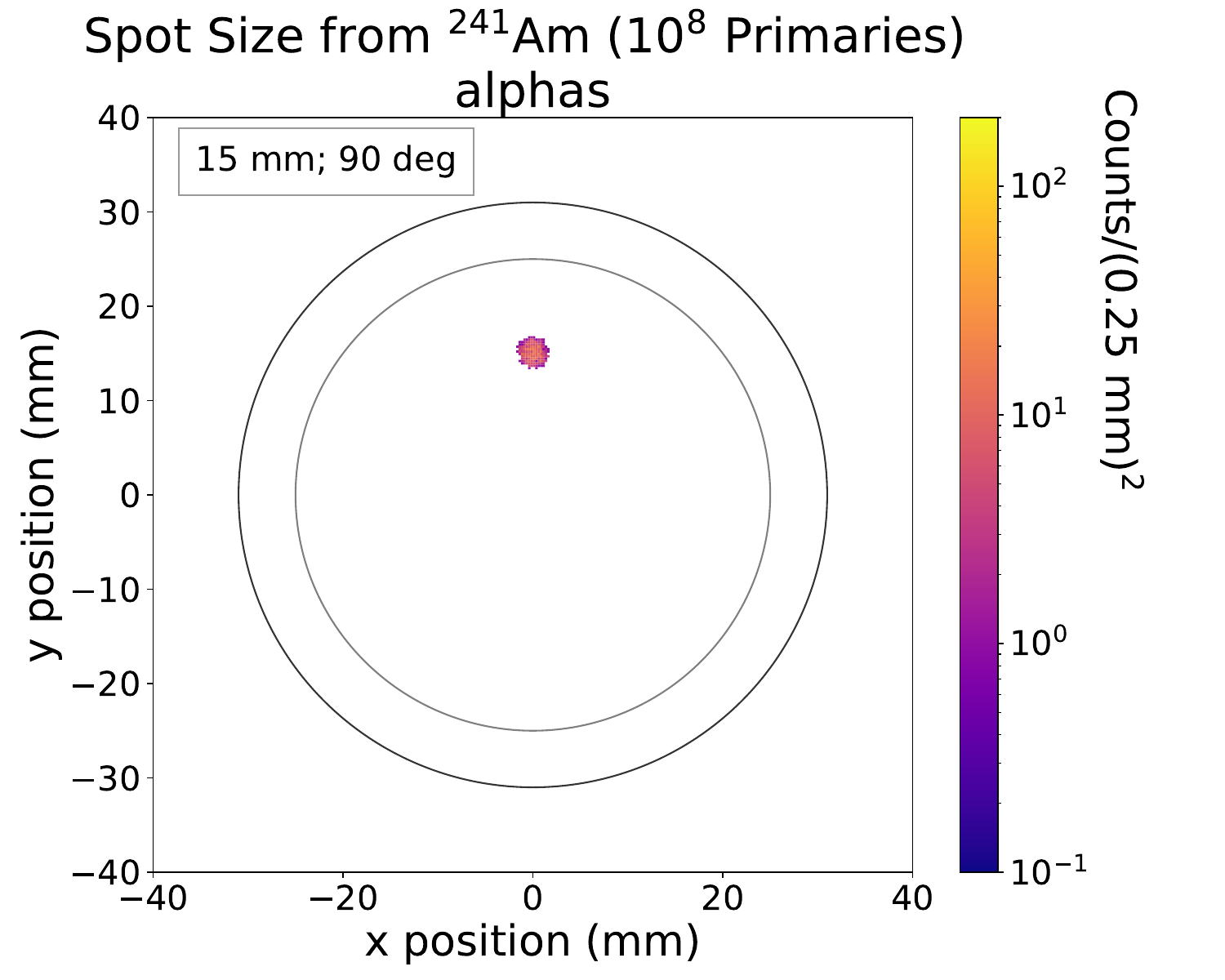}
    \caption{\label{subfig:2d_alphas_90deg}}
  \end{subfigure}
  \hfill
  \begin{subfigure}[t]{0.45\textwidth}
    \centering
      \includegraphics[scale=0.30]{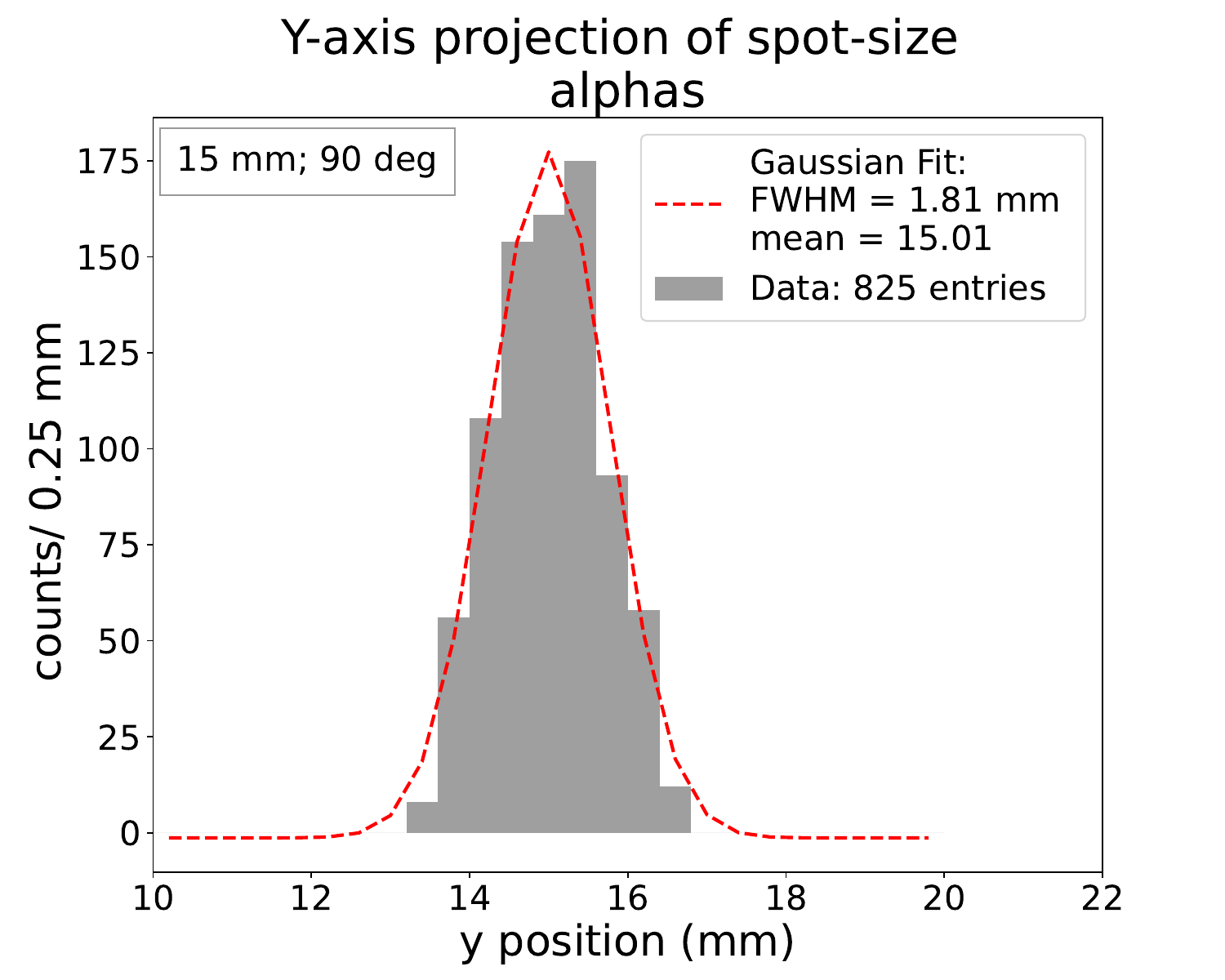}
    \caption{\label{subfig:1d_alphas_90deg}}
  \end{subfigure}
\caption{Simulated two-dimensional histograms of the X and Y interaction positions of the beam particles on the detector surface and their one-dimensional projections, for the collimator at normal incidence. The light grey circular region represents the edge of the detector surface where the bevel begins, and the dark grey represents the outer diameter of the detector (see Fig.~\ref{fig:collimator}).  (a, b) all particle interactions resulting from the $^{241}$Am source, including secondaries; (c, d) only 60~keV gamma particle interactions; (e, f) only alpha particle interactions.}
\label{fig:2D_spot}
\end{center}
\end{figure}

In order to estimate the spot-size of the $^{241}$Am source using the collimator above, we performed Monte-Carlo simulations again using g4simple. In our simulations, and later in the commissioning described in Sec.~\ref{sec:60keV}, we use the ``OPPI-1'' detector. OPPI-1 is an R\&D PPC detector made by ORTEC~\cite{ortec} from $^{\textrm{nat}}$Ge, shown in Fig. \ref{fig:front-end} installed on the CAGE cold plate. The detector features a 50~mm passivated surface, with 45\textdegree\ bevel and 62~mm total outer diameter. In a simulation with the OPPI-1 geometry, depicted in Fig.~\ref{fig:oppi_sims_vis}, we simulated 10$^8$ primary decays of $^{241}$Am for each simulated position, using a realistic implementation of the source disk within the collimator. For our 40~kBq source, $10^8$ decays represent about 42 minutes of real-time, background-free data. We performed simulations with the collimator at normal incidence with respect to the detector surface, and with the collimator rotated up to 45\textdegree, but incident on the same radial position on the detector surface. The resulting size on the detector surface with the source at normal incidence can be seen in the histograms in Fig.~\ref{fig:2D_spot}.
\clearpage
\clearpage

\begin{figure}
    \centering
    \includegraphics[width=0.5\textwidth]{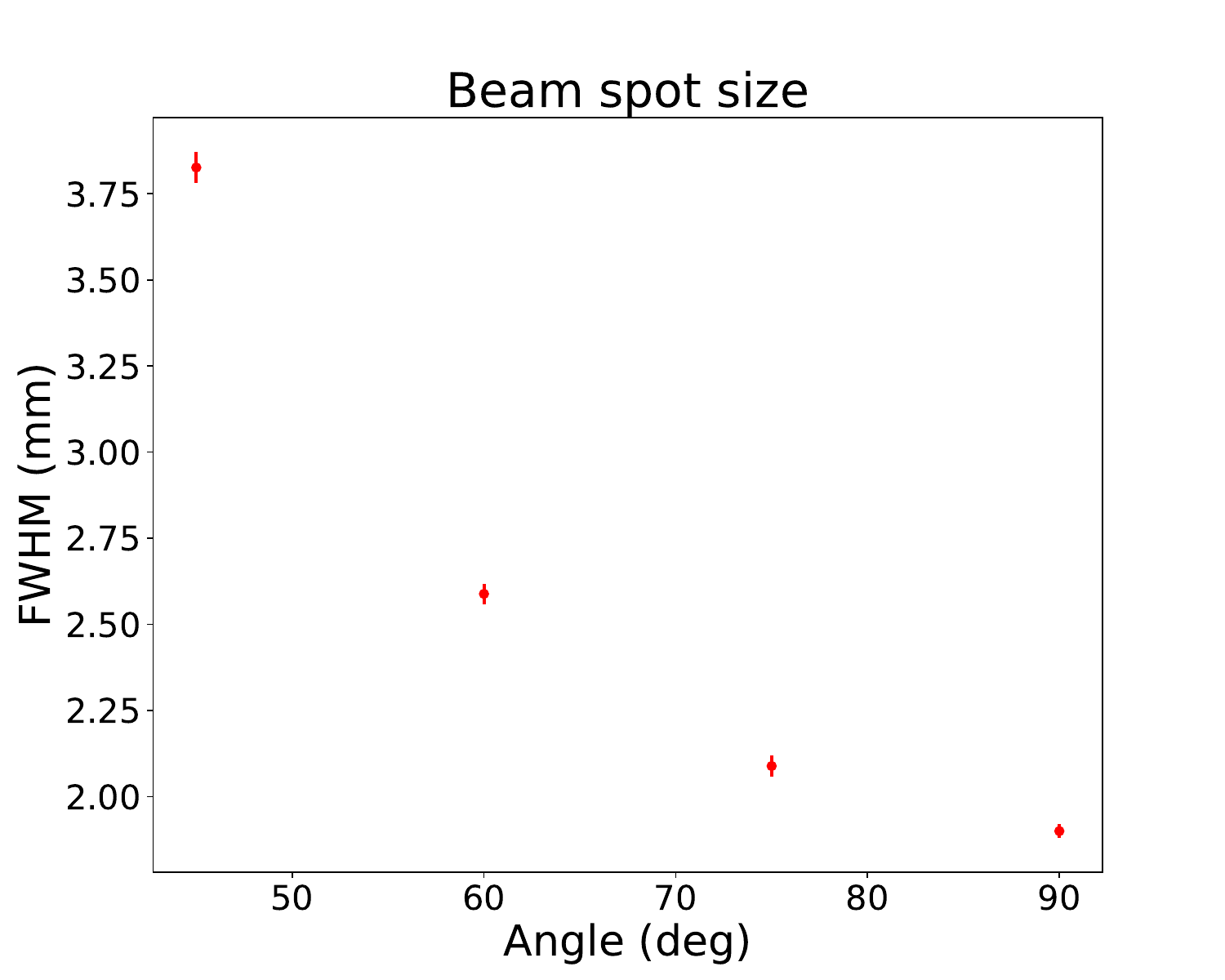}
    \caption{Spot size as a function of source angle for all particles interactions. The errorbars represent the uncertainty in the FWHM from the fit. Normal incidence corresponds to 90$^\circ$, as defined in Fig.~\ref{subfig:oppi}.}
    \label{fig:spot-size}
\end{figure}

To determine the beam spot-size from our simulations, we project the data onto the axis with the most beam spread (here the Y axis). We fit the resulting one-dimensional position data with a Gaussian function and use the FWHM from the fit as a measure of the beam spot-size.  Due to geometric effects, the beam spot becomes elongated with higher rotation angles of the source. This can be seen clearly in Fig.~\ref{fig:spot-size} showing the FWHM of the resulting Gaussian fit with respect to various rotation angles. 

\section{Commissioning Results: Analysis of surface events from 60 keV gammas}
\label{sec:60keV}

To commission the CAGE test stand we performed basic linear scans with our collimated $^{241}$Am source (see Sec.~\ref{sec:cage_collim}), and studied the associated 60~keV gamma. In the commissioning campaign, we again utilized the OPPI-1 detector, described in \ref{sec:sims}, and shown in Fig.~\ref{fig:holder} installed on the CAGE cold plate, with dimensions given in Fig.~\ref{subfig:oppi}. In this data campaign, the collimator remained at normal incidence with respect to the passivated surface of the detector and was translated to varying radial positions on the detector surface using only the linear motor stage.  Measurements were taken with the source positioned at 2.5~mm to 22.5~mm from the center of the detector in 5~mm increments, at the same rotary angle. During these scans, we estimate a $\sim$1~mm uncertainty in radial position, due to the systematic uncertainty of the collimator installed on the IR shield.

We utilized the Pygama \cite{pygama} Python-based analysis software developed by the LEGEND collaboration to analyze the resulting data. We calibrated the data by first applying trapezoidal filters to the raw waveforms, and picking off a fixed-timepoint from the filtered waveform~\cite{Majorana:2022vai}. This results in a parameter for energy that is more robust against noise than simply using the waveform maximum. We then make a correspondence to known peaks such as $^{40}$K, $^{208}$Tl, and $^{212}$Pb. Next, we applied basic data-cleaning cuts to remove background events, for example from electronics noise, pile-ups, and saturated events from cosmic muons. We used cuts on the mean and standard deviation of the baseline (first 3500 samples) of the waveform to remove events with excessive electronics noise, and cuts on the baseline slope to remove late pile-up events. We removed other pile-up events and muons using a ratio between the waveform maximum and the calibrated energy. After data-cleaning cuts, we used an energy-based cut to select 60~keV gammas for the following analyses.

\subsection{Rate Determination}\label{subsection:60keV-rate}

\begin{figure}[h!]
% \begin{center}
    \begin{subfigure}[t]{0.45\textwidth}
      \centering
        \includegraphics[scale=0.45]{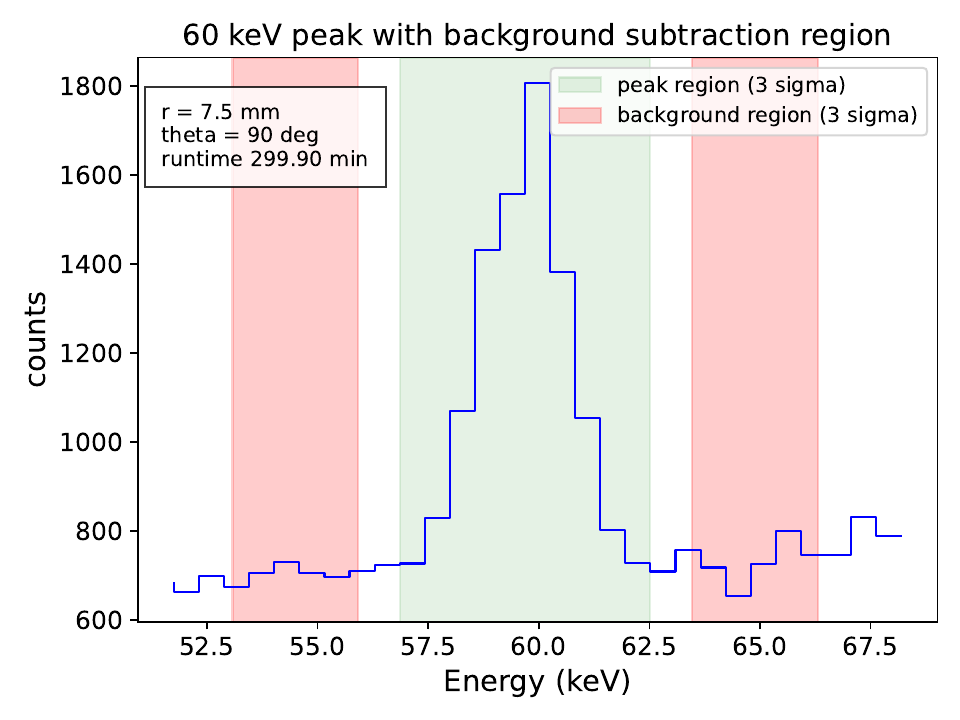}
      \caption{\label{subfig:backgroundSub}}
    \end{subfigure}
    \hfill
    \begin{subfigure}[t]{0.45\textwidth}
      \centering
        \includegraphics[scale=0.45]{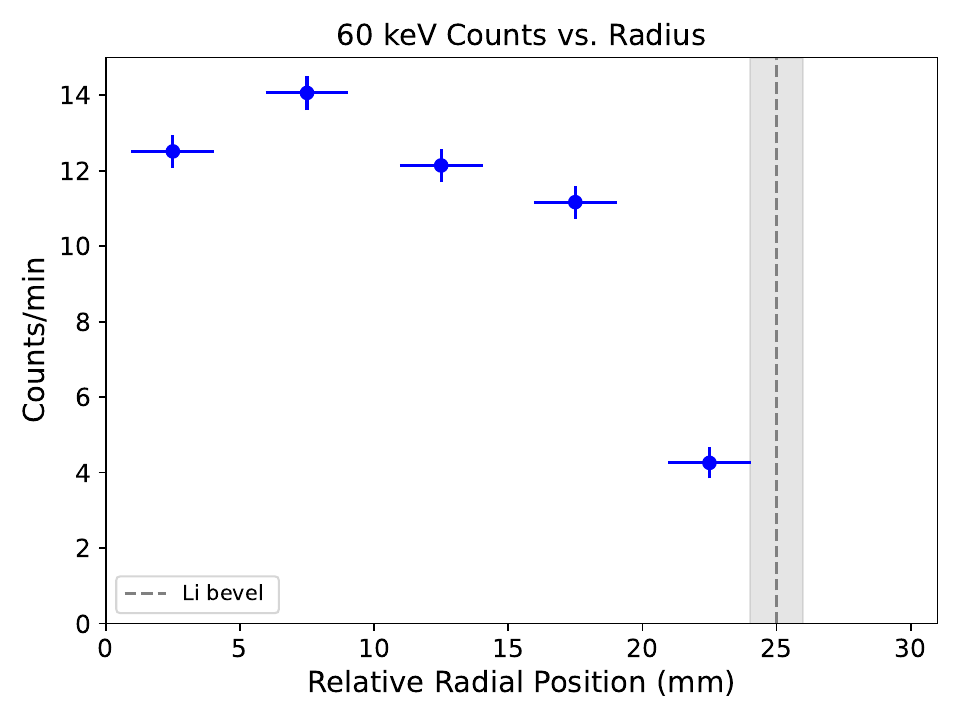}
      \caption{\label{subfig:60keV-rate}}
  \end{subfigure}
\caption{(a) Demonstration of the selection of signal and background events based on energy for the 60~keV rate determination. This example was at the 7.5~mm scan position. (b) Rate versus radial position for all scans. The errorbars in the rate are based on Poissonian errors. The errorbars in the radial position do not represent the uncertainty of the center of the beam, but the range of radial positions probed in each scan. They are the 2~mm FWHM beam-spot size added in quadrature with the 1~mm (2.35~mm FWHM) relative uncertainty of the  motor positions, as described in Sec.~\ref{sec:cage_motor}, which gives a value of 3.1~mm FWHM. The dotted line indicates the expected position of the Li bevel (N+ contact), and the shaded area represents the $\sim$1~mm thickness of the Li layer.
\label{fig:60keV-rate}}
% \end{center}
\end{figure}

 To determine the rate of 60~keV gammas, we counted events within 3$\sigma$ of the 60~keV peak and subtracted background counts determined from the sidebands of the peak. We chose sidebands of equal width to the peak window, from -5.5 to -4$\sigma$ on the left of the peak, and from 4 to 5.5$\sigma$ on the right of the peak. This is demonstrated in Fig.~\ref{subfig:backgroundSub} for a scan position at 7.5~mm from the detector center. We show the resulting background-subtracted rates for each scan position in Fig.~\ref{subfig:60keV-rate}. We see that the rate of 60~keV events drops off at about 22.5~mm. This is consistent with the relative $\sim$1~mm uncertainty in the scan position (2.35~mm FWHM) due to the motors, coupled with the unknown uncertainty in the absolute position, as discussed in Sec.~\ref{sec:cage_motor}, and with the 2~mm spot-size, since part of the beam may be shining on the Li bevel of the detector, which is $\sim$1~mm thick.

\subsection{Waveform Analysis using Superpulses}

The primary goal of the CAGE test stand is to study the pulse shapes of events at various positions on the detector surface. In addition to energy-based analyses, we directly studied the pulses from the 60~keV gammas in this data campaign. Similar to the background subtraction performed in \ref{subsection:60keV-rate}, we can also perform a background subtraction when looking at average pulses, or superpulses. We call the total superpulse of all events within the 2$\sigma$ peak $S_{p}$, the average background pulse from the sidebands $S_{b}$, and the average pulse due to ``signal'' 60~keV gammas at the position of interest $S_{s}$, given by 

\begin{equation}\label{eq:sp_bkgSub}
S_{s} = \frac{C_p S_p - C_b S_b}{C_p-C_b}
\end{equation}

\noindent where $C_{p}$ and $C_{b}$ are the event counts in the peak and sideband regions, respectively.

For this study, we use 2$\sigma$ peak and sideband regions to determine $S_p$ and $S_b$, with the sideband regions for $S_b$ from -5 to -4$\sigma$ on the left side of the peak region and 4 to 5$\sigma$ to the right. Compared to the 3$\sigma$, region used in the event-rate analysis described in the previous section, this results in a purer source of 60~keV events, while still significantly reducing the noise compared to studying individual waveforms. 

\begin{figure}[h!]
  \begin{subfigure}[t]{0.45\textwidth}
    \centering
      \includegraphics[scale=0.22]{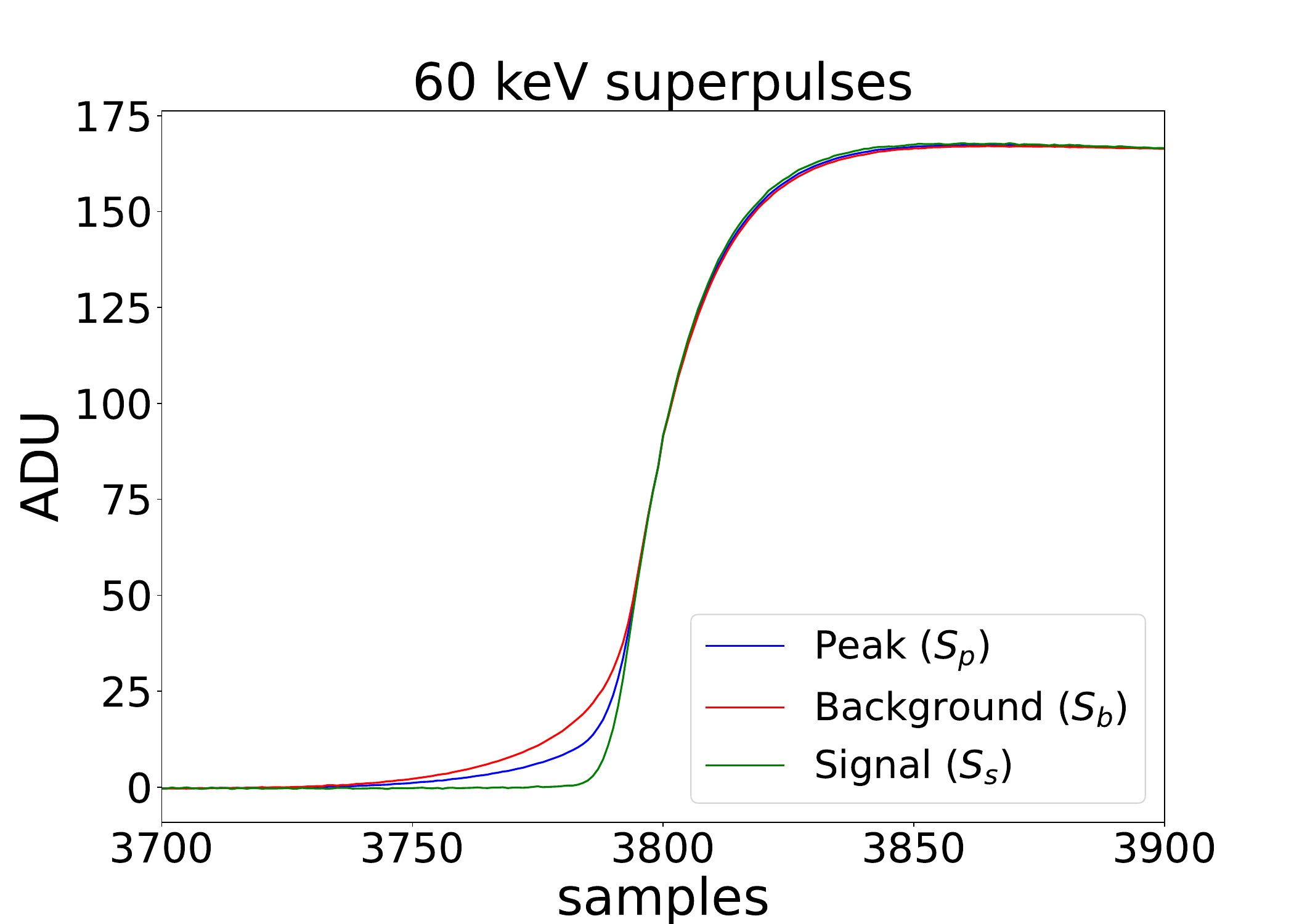}
    \caption{\label{subfig:sp_bkgSub_Notch}}
  \end{subfigure}
  \hfill
  \begin{subfigure}[t]{0.45\textwidth}
    \centering
      \includegraphics[scale=0.22]{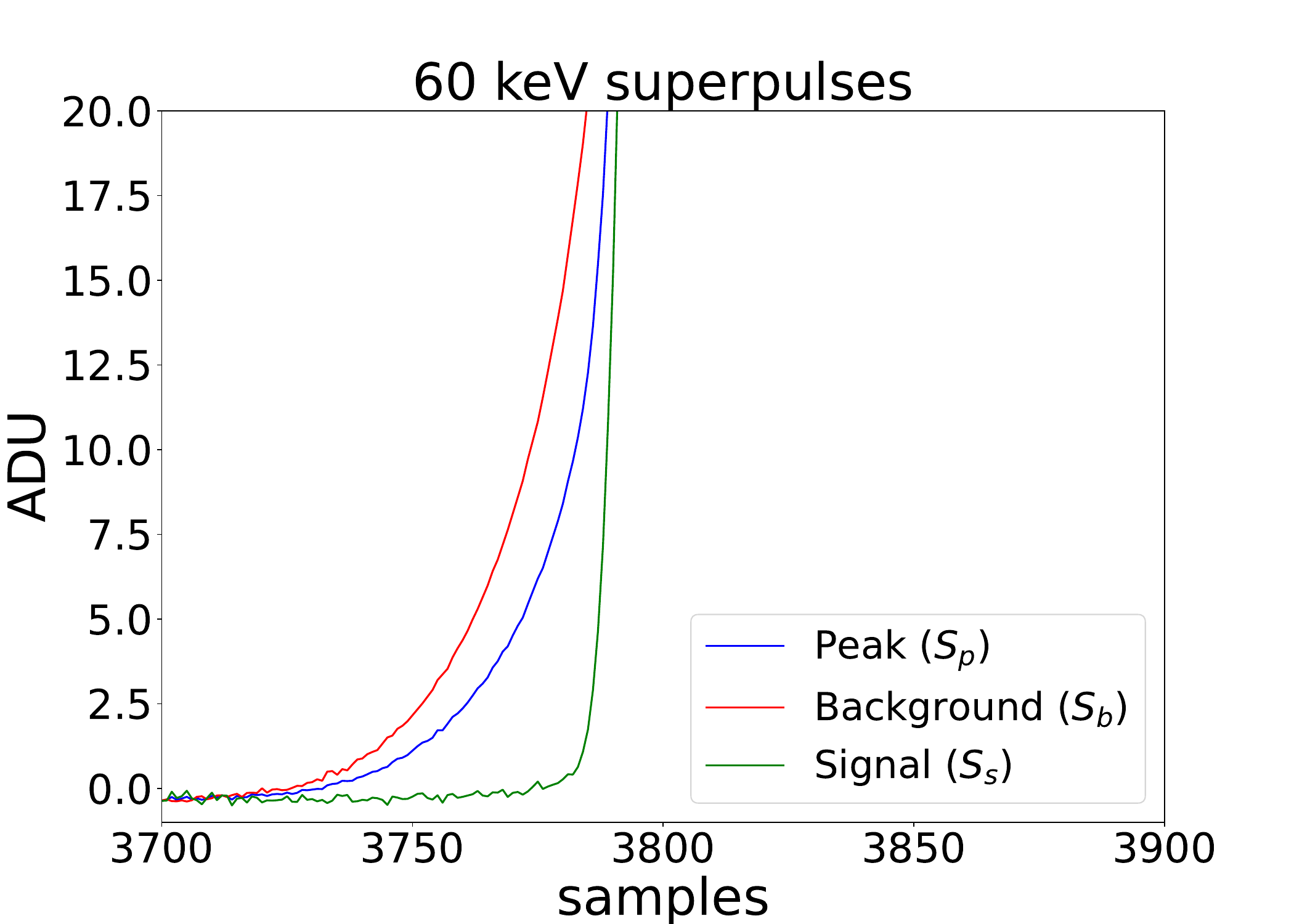}
    \caption{\label{subfig:sp_bkgSub_Notch}}
  \end{subfigure}
  \hfill
  \begin{subfigure}[t]{0.45\textwidth}
    \centering
      \includegraphics[scale=0.22]{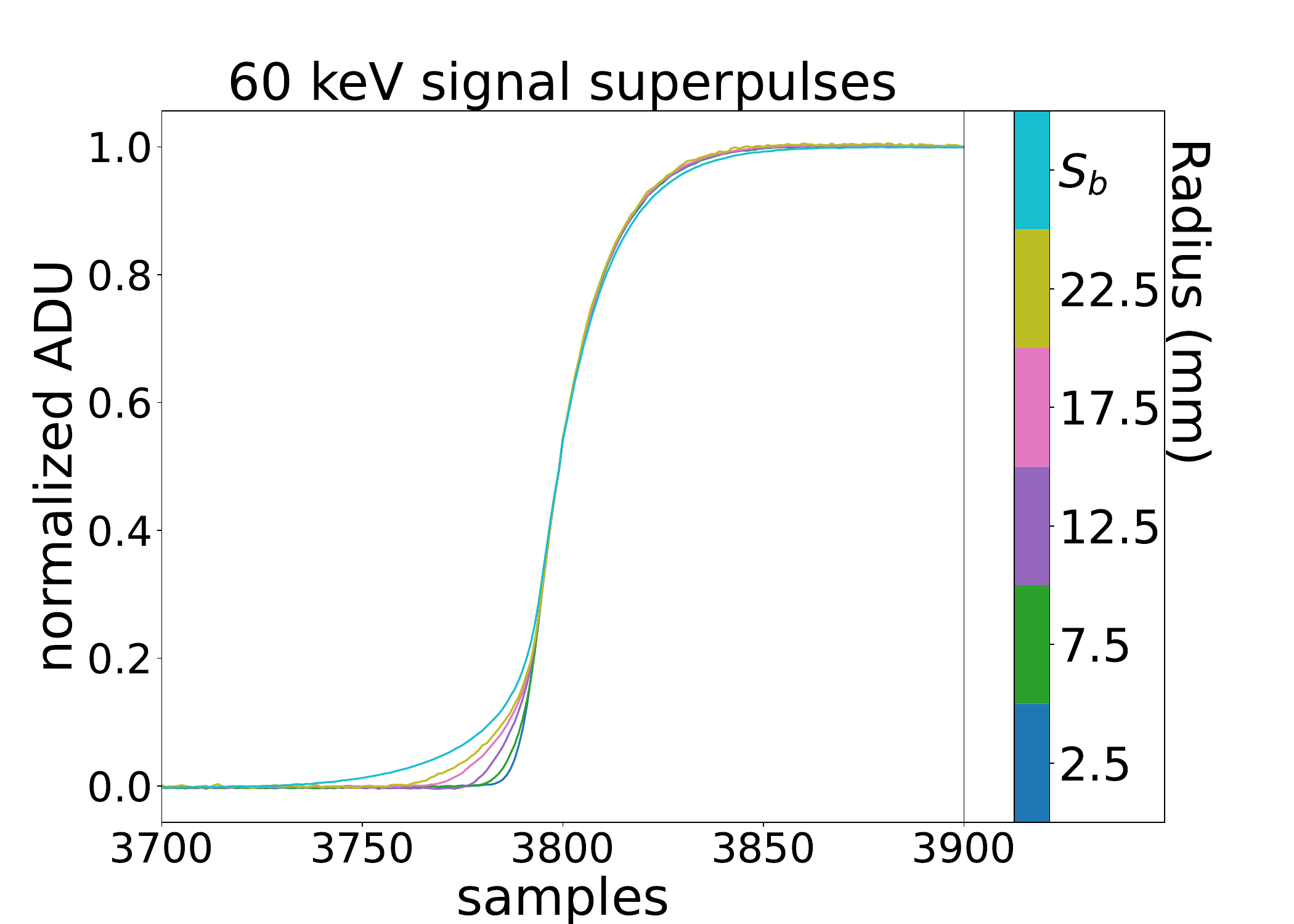}
    \caption{\label{subfig:super_riseTail}}
    \end{subfigure}
    \hfill
  \begin{subfigure}[t]{0.45\textwidth}
    \centering
      \includegraphics[scale=0.22]{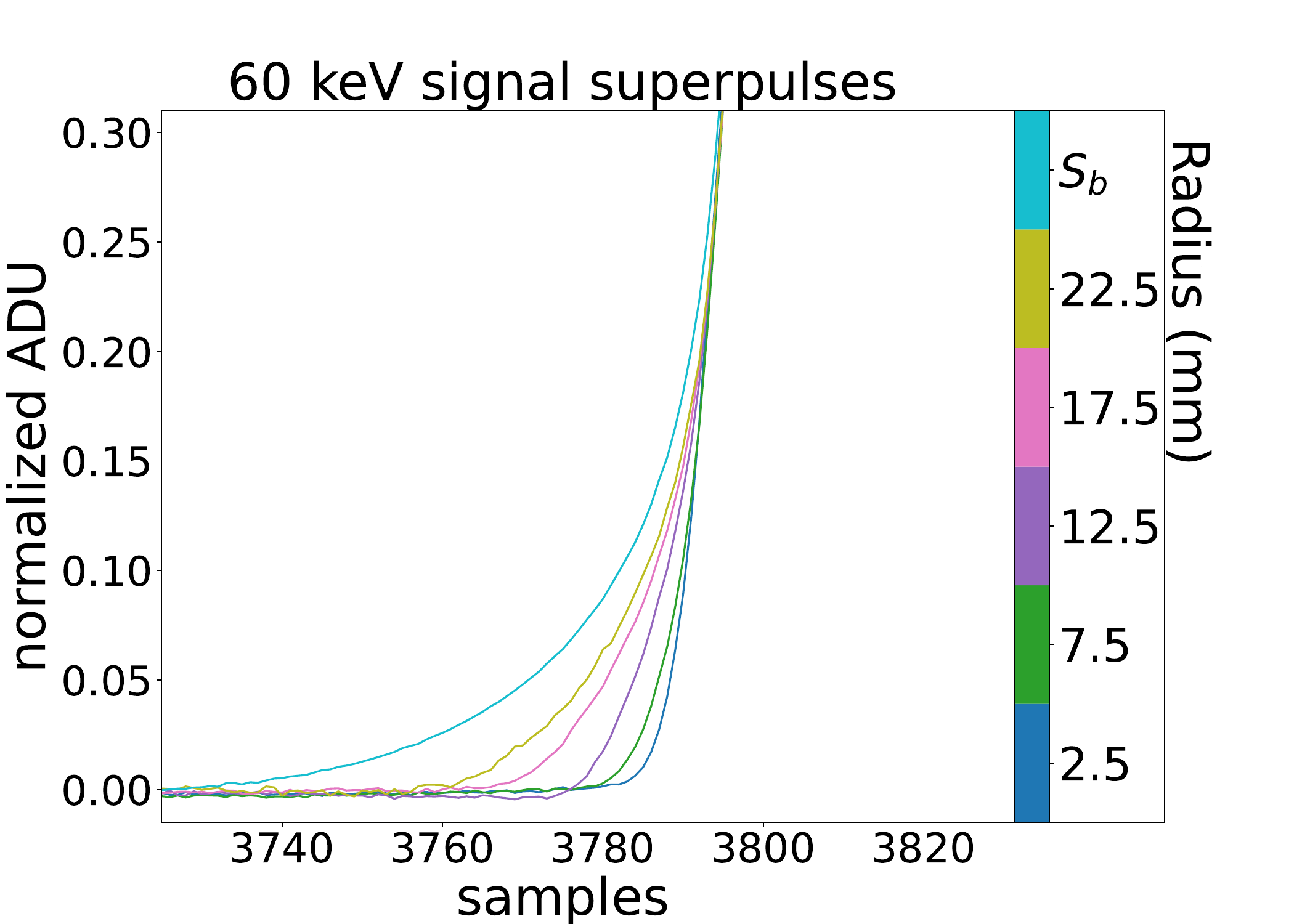}
    \caption{\label{subfig:super_rise}}
    \end{subfigure}
\caption{Top: the $S_p$, $S_b$, and $S_{s}$ superpulses from the 7.5~mm scan position. Bottom: normalized superpulses from the~60 keV gamma signal ($S_{s}$) for each scan position, including a superpulse from the sideband region (labeled ``$S_b$'').  Left plots show a 2 $\mu$s (200 sample) window around the waveform rising edge, with a zoom-in on the rise start on the right.
\label{fig:superpulses}}
\end{figure}

After selecting the appropriate peak and background regions, we calculate the superpulse waveforms $S_p$ and $S_b$ by averaging the waveforms in the peak and sideband regions, respectively. The waveforms are then baseline subtracted so that the waveform baseline has an average at 0 ADC, following which we apply a double pole zero (PZ) correction to correct for the affect of the RC decay from the electronics on the tail of the waveform. The resulting waveforms are also filtered from 50 and 25~MHz noise with a notch filter to remove polling noise. We then use Eq. \ref{eq:sp_bkgSub} to determine the superpulse due to 60~keV events, $S_{s}$. An example of $S_{p}$, $S_{b}$, and the resulting $S_{s}$ is shown in Fig.~\ref{fig:superpulses}(a,b). We repeat this process for each scan position, and the results of the superpulses from each scan position are shown in Fig.~\ref{fig:superpulses}(c,d). 

\begin{figure}[h!]
    \centering
    \includegraphics[scale=0.6]{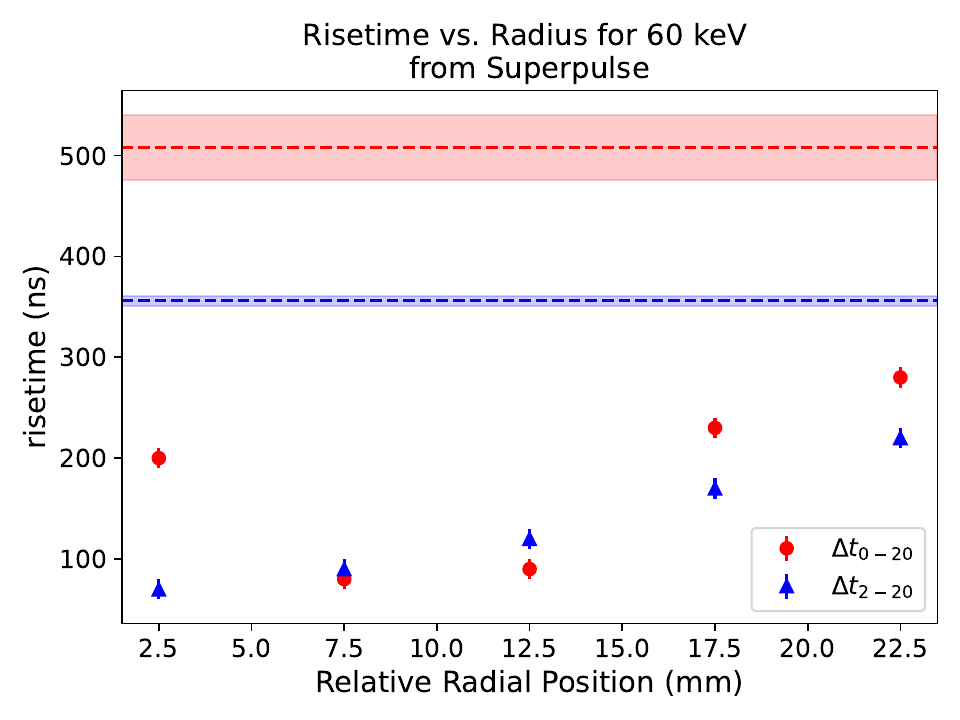}

\caption{Risetimes of the 60 keV superpulses, $S_{s}$, for each scan position. The lines and shaded regions indicate the average and standard deviation of the risetimes of the background superpulses, $S_b$, over all radii. The instability of the $t_0$ determination is clear from the larger standard deviation of $\Delta \textrm{t}_{0-20}$ of the background superpulses.} 
\label{fig:rt_superpulses}
\end{figure}

From Fig.~\ref{fig:superpulses}, it is clear that for 60~keV events, the risetimes of the signal superpulses $S_{s}$ depend on the radial position on the detector surface and are faster than the background superpulses $S_b$ in each position. We verified that the background superpulses ($S_b$) are also very consistent across scan positions, and represent a homogenous distribution of events throughout the bulk of the detector. We can exploit these risetime features for background reduction of surface events in the LEGEND experiment. Since the superpulses show the most deviation between 0 and 20\% of the waveform maximum, we focus on risetimes between 0 and 20\% max, $\Delta \textrm{t}_{0-20}$, and between 2 and 20\%, $\Delta \textrm{t}_{2-20}$. We determine $t_0$ from an asymmetric trapezoidal filter with a 20~ns integration time and 1~$\mu$s flat-top time. We define $t_0$ as the last timepoint prior to the maximum of the asymmetric trapezoid at which the output passes 0 ADC. To determine the 20\% (2\%) timepoint, we calculate 20\% (2\%) of the maximum value of the superpulse and determine the first instance in time that value was reached. This value is very stable since we are determining it from superpulses with little noise. The risetimes for each position are shown in Fig.~\ref{fig:rt_superpulses} compared to the average risetime of the background superpulse ($S_b$). In choosing a risetime parameter for event discrimination, we would look for one that has the largest deviation across radial positions, and has maximum separation from the background value. Though on first glance this would appear to be $\Delta \textrm{t}_{0-20}$, we found the $t_0$ calculation unstable, especially at lower radii. This is also visible in Fig.\ref{fig:rt_superpulses} from the wider standard deviation of $\Delta \textrm{t}_{0-20}$ for the background superpulses. We therefore recommend that the $\Delta \textrm{t}_{2-20}$ parameter be further explored by LEGEND.   

\subsection{Alpha Event Discrimination}

Although our first commissioning run focused on the analysis of 60~keV gamma events from $^{241}$Am, we also successfully identified alpha events from the same scan. We accomplished this using the delayed charge recovery, or DCR, parameter developed by the \textsc{Majorana} collaboration to identify and remove backgrounds from surface alphas \cite{dcr}. There is energy degradation of alpha events on the passivated surfaces of PPC detectors due to charge trapping. The DCR parameter measures the release of delayed, trapped charges from alpha interactions, which are apparent on the tail of pole-zero corrected waveforms as a positive slope. As implemented in our analysis, we apply a trapezoidal filter with a integration time of 7.5~$\mu$s and a flat-top of 22.5 $\mu$s to the pole-zero corrected waveforms. We choose a pickoff value from the resulting trapezoid at 79~$\mu$s, near the end of the trace, which we call DCR. In this way, we are effectively calculating an average slope of the tail between 56.5 (79 - 22.5) and 79~$\mu$s, in a more computationally efficient way than fitting the tail of the pole-zero corrected waveform. 

\begin{figure}[h!]
  \begin{subfigure}[t]{0.5\textwidth}
    \centering
      \includegraphics[scale=0.45]{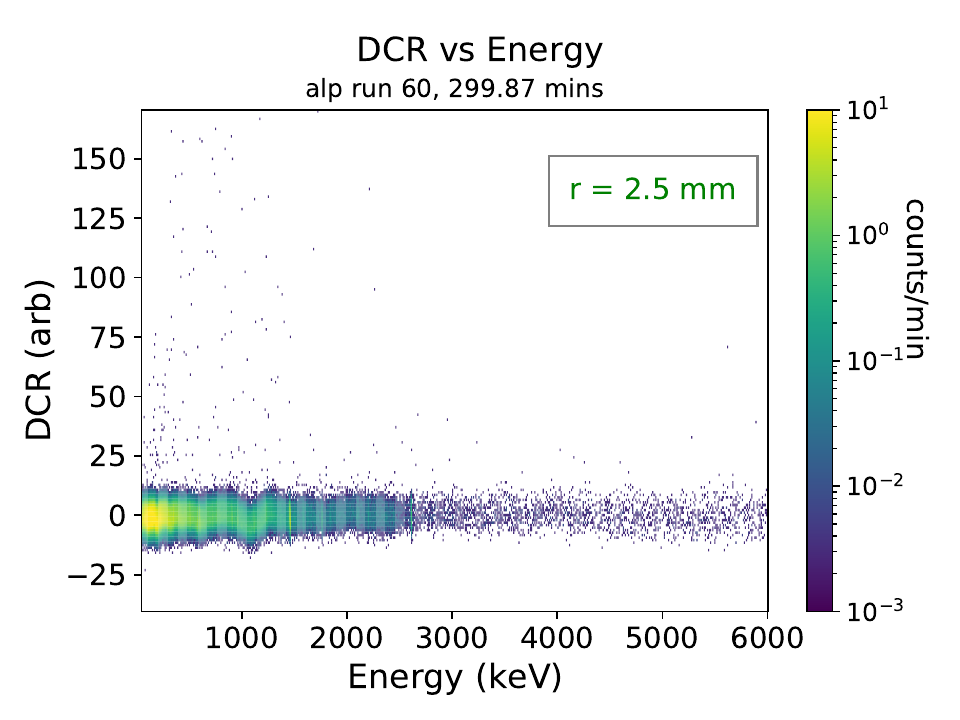}
    \caption{\label{subfig:dcr_2.5}}
  \end{subfigure}
  \hfill
  \begin{subfigure}[t]{0.45\textwidth}
    \centering
      \includegraphics[scale=0.45]{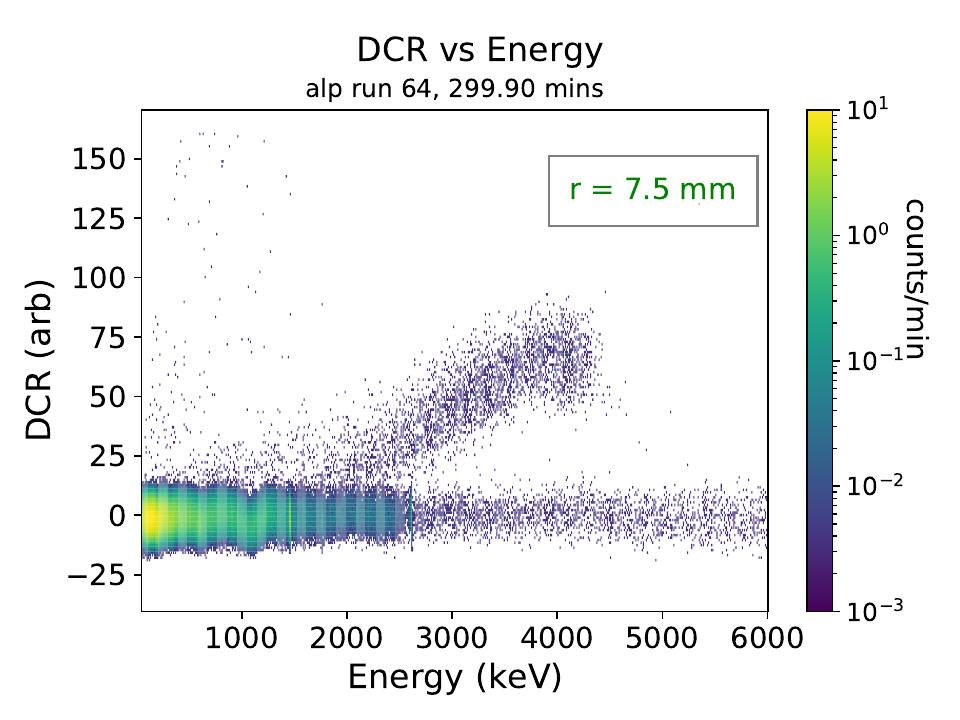}
    \caption{\label{subfig:dcr_7.5}}
  \end{subfigure}
  \hfill
  \begin{subfigure}[t]{0.45\textwidth}
    \centering
      \includegraphics[scale=0.45]{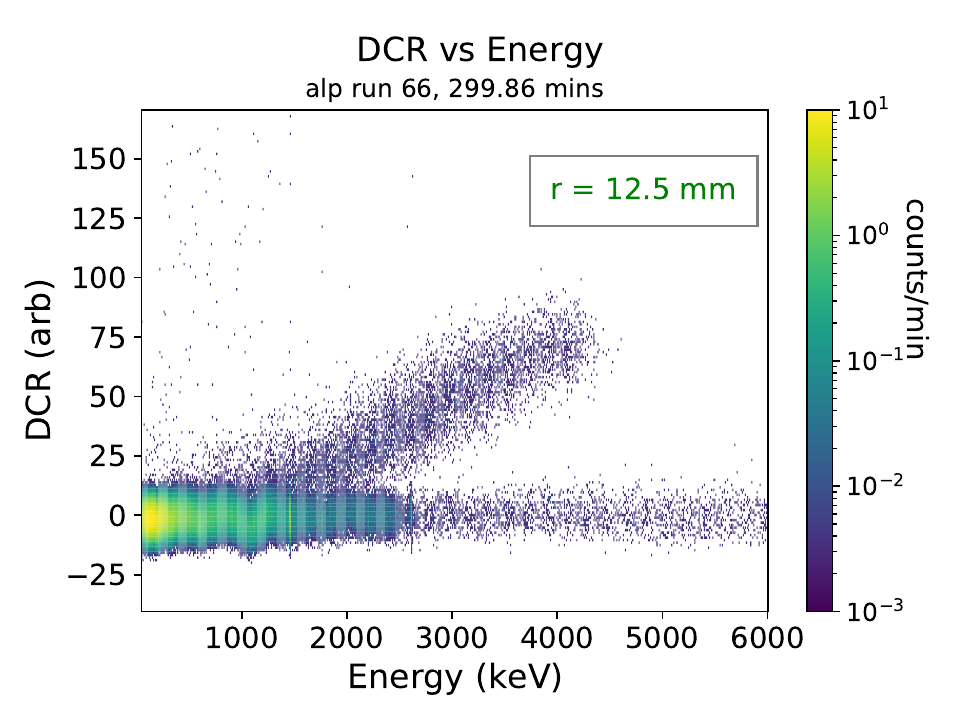}
    \caption{\label{subfig:dcr_12.5}}
  \end{subfigure}
  \hfill
  \begin{subfigure}[t]{0.45\textwidth}
    \centering
      \includegraphics[scale=0.45]{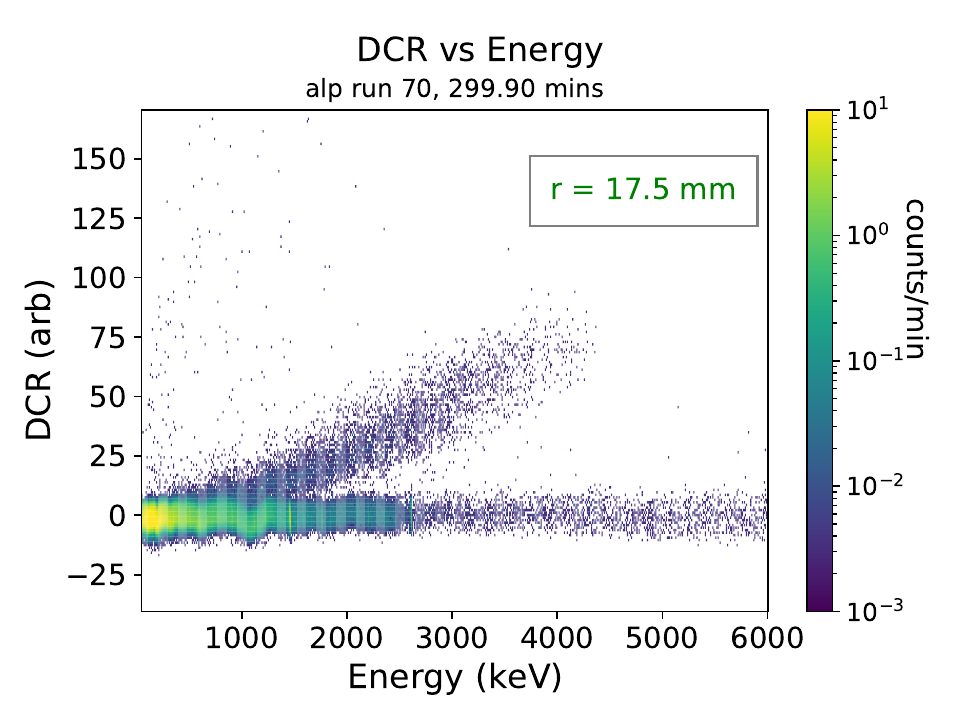}
    \caption{\label{subfig:dcr_17.5}}
  \end{subfigure}
  \hfill
  \begin{subfigure}[t]{0.45\textwidth}
    \centering
      \includegraphics[scale=0.45]{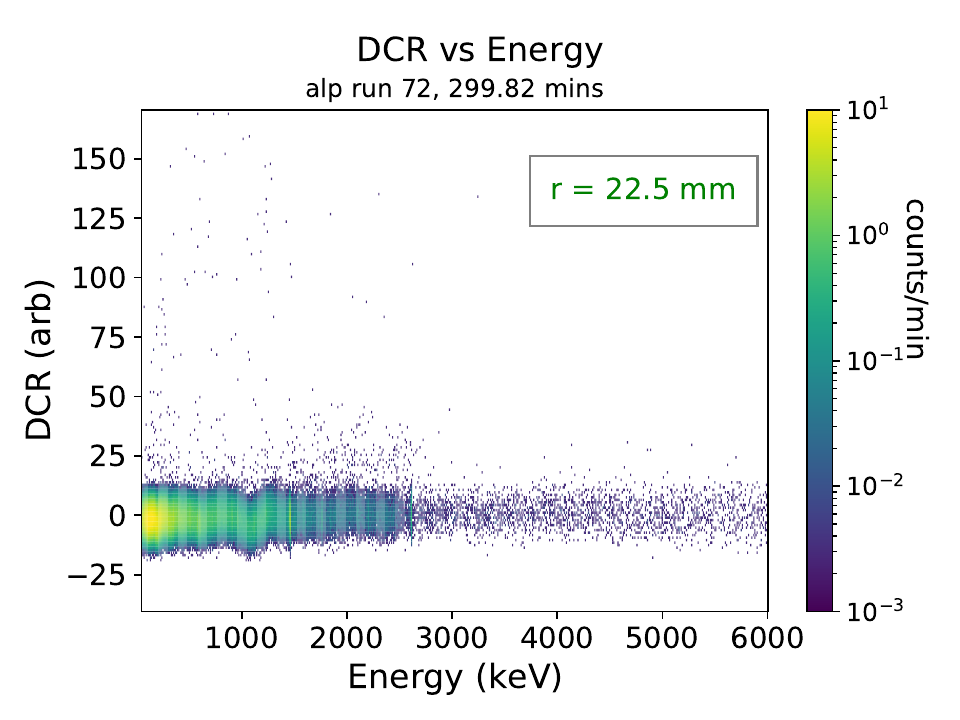}
    \caption{\label{subfig:dcr_22.5}}
  \end{subfigure}
\caption{Two-dimensional histogram of the DCR parameter plotted against calibrated energy for beam positions of (a) 2.5 mm, (b) 7.5 mm, (c) 12.5 mm, (d) 17.5 mm, and (e) 22.5 mm. Alphas are visible at higher values of DCR, above about 20. We attribute the small oscillation in the DCR parameter when compared against energy to ADC non-linearities~\cite{MAJORANA:2020llj}.
\label{fig:dcr}}
\end{figure}

Fig.~\ref{fig:dcr} shows the DCR values for all linear scan positions as a function of the event energy. The bulk events appear distributed around 0 DCR across all energies. There are clear populations of energy-degraded alphas seen at higher values of DCR that develop at the 7.5~mm scan position, intensify, then slowly disappear by the 22.5~mm scan positions. These results are consistent with alphas from the source beam being blocked by the PEEK diving board holding the front-end electronics at low radii, and the detector holder ring as well as the Li N+ contact at high radii. The energy of the original alphas from $^{241}$Am is 4.5~MeV, and the apparent smearing in DCR over the energy ranges is due to energy degradation of alphas at the surface of the detectors.

\section{Conclusions and Future Work}
\label{conclusions}

We have built and demonstrated the functionality and stability of a new scanning system to study surface events on PC detectors. Our scanning system offers improvements over previous systems by incorporating a rotatable source mounted internal to movable IR shield. With three movement stages, the source can be positioned freely at any point on the passivated surface of the detector, and the angle of incidence of the beam with respect to the surface can be changed to study effects at varying depths on the detector face. 

We have shown that we are able to detect 60~keV gammas as well as alphas from $^{241}$Am with good separation from bulk events in the detector. In particular, we have shown that 60 keV gammas are characterized by radially-dependent risetimes, and we have recommended concrete risetime parameters with which the LEGEND experiment can identify surface events even at low-energy. This study is a good example of the power of using PC detectors for fundamental physics searches. 

In future work, we will study the energy degradation of alphas on the surfaces of various PC detectors with CAGE in dedicated scans, with the goal of developing more robust models of the surface alpha interactions and more effectively identifying and removing those populations in LEGEND data. Here, it will be of particular interest to change the angle of incidence of the source beam with respect to the detector surface. This will allow the study of alphas at varying depths within the detector surface, at the same radial position. This has not been studied before to our knowledge, and will enable more detailed model building of alpha events on PC detector surfaces, which can lead to an even better discrimination power of alpha backgrounds in the LEGEND experiment.

%__________
% Note on appendices from JINST:
% \appendix
% \section{Some title}
% Please always give a title also for appendices.
%__________

%__________
% Note on acknowledgments from JINST:
% \acknowledgments
%
% This is the most common positions for acknowledgments. A macro is
% available to maintain the same layout and spelling of the heading.
%
% \paragraph{Note added.} This is also a good position for notes added
% after the paper has been written.
%__________
\section*{Acknowledgments} 
We would like to sincerely thank Steve Elliott for loaning us the OPPI-1 detector that was used in these scans. We thank Julieta Gruszko and Sebastian Alvis for their work on the preliminary design of CAGE, as well as for many helpful conversations about CAGE's design and construction. We would also like to thank Matthew Busch for offering very insightful technical advice during the construction phase. Last but not least, we thank Iris Abt, Oliver Schulz, and the group at the Max Planck Institute for Physics for very useful feedback and conversations. This material is based upon work supported by the U.S. Department of Energy, Office of Science, Office of Nuclear Physics under Award Numbers DE-FG02-97ER41041, DE-FG02-97ER41033, and the National Science Foundation.

%__________
%Note on References from JINST:
% \begin{thebibliography}{99}
%
% \bibitem{a}
% Author, \emph{Title}, \emph{J. Abbrev.} {\bf vol} (year) pg.
%
% \bibitem{b}
% Author, \emph{Title},
% arxiv:1234.5678.
%
% \bibitem{c}
% Author, \emph{Title},
% Publisher (year).
% Also, please have only one work for each \bibitem.
% \end{thebibliography}
%__________

\bibliographystyle{unsrt}
\bibliography{cage_inst}
\end{document}